\documentclass[final,twocolumn,10pt,letterpaper]{IEEEtran}

\usepackage{cite}
\ifCLASSINFOpdf
   \usepackage[pdftex]{graphicx}
  \DeclareGraphicsExtensions{.pdf,.jpeg,.png}
\else
   \usepackage[dvips]{graphicx}
   \DeclareGraphicsExtensions{.eps}
\fi

\usepackage{hyperref}
\usepackage[cmex10]{amsmath}
\usepackage{amssymb,amsthm}
\usepackage{dsfont}

\newtheorem{remark}{Remark}
\newtheorem{lemma}{Lemma}
\newtheorem{theorem}{Theorem}
\newtheorem{corollary}{Corollary}

\newtheorem{proposition}{Proposition}

\usepackage{psfrag}
\usepackage{array}
\usepackage{upgreek}
\usepackage{mdwmath}
\usepackage{mdwtab}
\usepackage{enumerate}
\usepackage{stfloats}

\newcommand{\mathdef}{~\raisebox{-0.03cm}{$\triangleq$}~}
\usepackage{mathtools}

\usepackage[caption=false,font=footnotesize,labelfont=sf,textfont=sf]{subfig}

\newcommand{\sig}{\mathsf{P}}
\newcommand{\inter}{\mathsf{\tilde{J}}}
\newcommand{\lap}{\mathcal{L}}

\begin{document}

%
\title{A Tractable Model for Non-Coherent Joint-Transmission Base Station Cooperation}

\author{Ralph Tanbourgi\IEEEauthorrefmark{1}, Sarabjot Singh\IEEEauthorrefmark{2}, Jeffrey G. Andrews\IEEEauthorrefmark{2}, and
Friedrich K. Jondral\IEEEauthorrefmark{1}
\thanks{\IEEEauthorrefmark{1}R.~Tanbourgi and F.~K.~Jondral are with the Communications Engineering Lab (CEL), Karlsruhe Institute of Technology (KIT), Germany. Email: \texttt{\{ralph.tanbourgi, friedrich.jondral\}@kit.edu}. This work was partially
supported by the German Academic Exchange Service (DAAD) and the German Research Foundation (DFG) within the Priority Program 1397 "COIN" under grant No. JO258/21-1.}
\thanks{\IEEEauthorrefmark{2}S.~Singh and J.~G.~Andrews are with the Wireless and Networking Communications Group (WNCG), The University of Texas at Austin, TX, USA. Email: \texttt{\{sarabjot, jandrews\}@ece.utexas.edu}}
}


\maketitle

\begin{abstract}
This paper presents a tractable model for analyzing non-coherent joint transmission base station (BS) cooperation, taking into account the irregular BS deployment typically encountered in practice. Besides cellular-network specific aspects such as BS density, channel fading, average path loss and interference, the model also captures relevant cooperation mechanisms including user-centric BS clustering and channel-dependent cooperation activation. The locations of all BSs are modeled by a Poisson point process. Using tools from stochastic geometry, the signal-to-interference-plus-noise ratio ($\mathtt{SINR}$) distribution with cooperation is precisely characterized in a generality-preserving form. The result is then applied to practical design problems of recent interest. We find that increasing the network-wide BS density improves the $\mathtt{SINR}$, while the gains increase with the path loss exponent. For pilot-based channel estimation, the average spectral efficiency saturates at cluster sizes of around $7$ BSs for typical values, irrespective of backhaul quality. Finally, it is shown that intra-cluster frequency reuse is favorable in moderately loaded cells with generous cooperation activation, while intra-cluster coordinated scheduling may be better in lightly loaded cells with conservative cooperation activation.
\end{abstract}


\IEEEpeerreviewmaketitle

\section{Introduction}\label{sec:introduction}
Base station (BS) cooperation---described varyingly as coordinated multi-point, network multiple-input multiple-output (MIMO) or more recently as a cloud radio access network (C-RAN)---has garnered significant research attention since it is a theoretical powerful method for ameliorating a key degradation in modern cellular systems: other-cell interference. In principle, BS cooperation mimics a large distributed MIMO system by letting a subset of BSs share their resources to jointly serve a subset of users \cite{3gpp_tr_36819,gesbert10,simeone12,sawahashi10}. Cooperation schemes may range from coordinated scheduling and beamforming to full joint-processing, depending on the employed backhaul architecture, tolerable mobility and complexity, and other constraints. Successful joint processing over a cluster of BSs can turn their interference back into useful signals \cite{sawahashi10,irmer11}, although the out-of-cluster interference still acts as noise \cite{ZhaChe09}. In non-coherent joint transmission (NC-JT), also called single-user JT, BSs cooperate by jointly transmitting the same data to a given user without prior phase mismatch correction and tight synchronization \cite{3gpp_tr_36819,docomo10,li12,ericsson11,barbieri12,lee12}. At the user, the resulting non-coherent sum of the useful signals yields a received power boost; an effect known as cyclic delay diversity for single-frequency networks using orthogonal frequency division multiplexing (OFDM)\cite{morimoto06,lee12,gosh10}. In some cases, NC-JT outperforms its coherent counterpart due to less stringent synchronization and channel state information (CSI) requirements\cite{li12}. NC-JT increases the cell load and is therefore considered for lightly-loaded cell scenarios \cite{ericsson11} only, where it can also be used for load-balancing purposes \cite{barbieri12}.

\subsection{Challenges facing BS Cooperation}
A prerequisite for most forms of BS cooperation is the availability of CSI at all the BSs of the same cooperative cluster. The CSI reported on the uplink to each BS is used to decide upon the users to be served jointly as well as on the resource allocation, possibly causing significant signaling overhead. To make things worse, this computational burden must be carried over finite-capacity backhaul links within a fraction of the channel coherence time. For joint transmission, all the user data has to be distributed among all cooperating BSs. Such considerations appear to prohibit large cooperative clusters, and one open question is the best cluster size for various types of cooperation. Even theoretically, it was recently shown that the benefits from cooperation are fundamentally limited \cite{lozano12}, and the gains obtained through larger clusters vanish beyond a certain size. This saturation point, in turn, obviously depends on many system aspects including radio channel, network geometry and interference; and finding that point inevitably requires to first disclose their complex interactions and understand their impact in a comprehensive way.

Because of these many complex interactions, studying BS cooperation is a challenge, requiring either simplistic models (Wyner or 2-cell) for use with analysis, or time-consuming system-level simulations where conclusions tend to be opaque in terms of the affect of the various simulation parameters. The lack of useful models for analyzing cooperative cellular networks was recently reported also by the authors of \cite{tukmanov13}, where the need for new tractable models for studying cooperative cellular networks was highlighted. The authors concluded that the tools provided by the \emph{stochastic geometry} framework \cite{stoyan95,HaenggiBook, tanbourgi13_1,tanbourgi13_2} may be the appropriate answer to the above shortcomings. Although promising, the application of stochastic geometry to the modeling of cooperative cellular networks entails some non-trivial challenges: (i) modeling the user-to-BS association is more difficult since a user may now be served (indirectly) by multiple BSs; (ii) since BS cooperation opportunistically exploits small-scale channel fluctuations\cite{gesbert10}, channel-dependent cooperation incentives should not be excluded from the model; (iii) for joint transmission, the received sum signal power and sum interference, both possibly originating from the same source of randomness, must be simultaneously characterized.

In this work, we address the above challenges and derive a tractable model for downlink BS cooperation with NC-JT. Using tools from stochastic geometry, we characterize the $\mathtt{SINR}$ distribution at a typical user in a generality-preserving way. This model, being the key contribution itself, shall provide both a more nuanced understanding of the system behavior as well as a useful tool for design purposes. The major contributions of this work are summarized in Section~\ref{sec:contributions}.

\subsection{Related Work}\label{sec:prior_work}
In \cite{dhillon12}, the authors derived the coverage probability for a typical user with instantaneous $\max$-$\mathtt{SINR}$ BS association in a multi-tier cellular environment. Such a scheme can be seen as a form of BS cooperation known as dynamic cell selection (DCS)  \cite{sawahashi10,marsch11} or transmission point selection (TPS) \cite{lee12}. Although not explicitly termed as BS cooperation, a related cellular concept was investigated in \cite{keeler13}. Here, the coverage probability of a typical user having multiple links (to the first $k$ strongest) was derived. Cooperative beamforming and scheduling (CB/CS) in cellular networks with irregular BS locations was studied in \cite{huang11,huang12} for the case of static BS clustering. The authors showed that the scaling of the outage probability exponent as the average number of cooperative BSs increases depends on the amount of scattering. In \cite{jung13}, the gain of CS BS cooperation was studied for the worst-case user, i.e., the cell-edge user, using a Poisson-Voronoi model \cite{stoyan95}. 

Models for joint transmission have recently been proposed in\cite{baccelli13_1,haenggi13_comp}. In\cite{baccelli13_1}, the authors study BS {\it coherent} JT with different geometry-based cooperation rules (based on 2-Voronoi diagrams) under Rayleigh fading. In contrast to the present work, only pair-wise cooperation is considered and opportunistic cooperation decisioning is not modeled. The results obtained in this work, about 17\% in average coverage, may differ significantly when assuming multiple cooperative BSs, timing/phase mismatch, and other fading distributions. In\cite{haenggi13_comp}, the performance of NC-JT in a heterogeneous cellular network was characterized under Rayleigh fading. The authors considered the cases where a user is jointly served by either its $K$ strongest BSs or by its strongest BS from each tier, which implies that the cooperative cluster may potentially span a large area.

From a modeling point of view, in this work we shall consider the case of multiple cooperating BSs as in\cite{haenggi13_comp}, however, except for that we shall consider a {\it fixed} cooperation area due to, e.g., limited backhaul complexity/costs. In contrast to\cite{haenggi13_comp,baccelli13_1}, we shall furthermore assume an arbitrary finite-moment fading distribution as well as a channel-dependent cooperation activation. A summary and discussion of our results is provided in the following section.

\subsection{Contributions and Outcomes}\label{sec:contributions}
The mathematical model as well as the considered cooperation scenario are explained in Section~\ref{sec:model}.

\textit{Characterization of the $\mathtt{SINR}$ distribution (main result):} In Section~\ref{sec:characterization}, we characterize the $\mathtt{SINR}$ distribution for a typical user when being served by multiple cooperating BSs under NC-JT. Thereby, interference is due to out-of-cluster BSs and (possibly) intra-cluster BSs serving other users on the same radio resources. Owing to our approach, the main result is given in a compact semi-closed form (involving derivatives of elementary functions). As an additional attribute, the result enjoys a high degree of generality. For instance, we do not have to rely on a particular fading distribution.

\textit{Key insights:} Observations following from the main results are: (i) the $\mathtt{SINR}$ gain obtained through cooperation increases with the path loss exponent; (ii) for a fixed geographic cooperation region, it was shown that increasing the network-wide density of BSs decreases $\mathtt{SINR}$ outage probability exponentially. Remarkably, this BS ``densification'' can be done at random, i.e., no careful site planning is required.
		
\textit{Effect of imperfect CSI:} Assuming pilot-based channel estimation with minimum mean square error (MMSE) criterion, it is shown in Section~\ref{sec:csi} that imperfect CSI becomes the performance-limiting factor as the cluster size increases, which is consistent with prior findings. For typical scenarios, the point at which increasing the cluster size is no longer beneficial for NC-JT in terms of average spectral efficiency is roughly around 7 BSs. This means that, even with a perfect backhaul, going beyond these values results in almost no performance gain. Also, since the spectral efficiency metric does not account for the cell load increase created by NC-JT, the above value should be seen as a rough upper bound on practically relevant cluster sizes.
	
\textit{Intra-cluster scheduling:} An important question related to NC-JT is whether BSs of a cooperative cluster not participating in an ongoing joint transmission should reuse the radio resources allocated to the joint transmission. In Section~\ref{sec:scheduling}, it is shown that intra-cluster frequency reuse should be employed in moderately loaded cells, in particular when channel-dependent cooperation activation is in favor of triggering a joint transmission. In this regime, load gains can be harvested without much worsening of the $\mathtt{SINR}$. When a conservative channel-dependent cooperation activation is chosen, intra-cluster coordinated scheduling may be more appropriate in lightly loaded cells since additional gains from muting intra-cluster interference can then be obtained.

\textbf{Notation:} 
Sans-serif letters ($\mathsf{z}$) denote random variables while serif letters ($z$) denote their realizations or variables. 

\begin{figure}[t]
	\centering
    \includegraphics[width=0.47\textwidth]{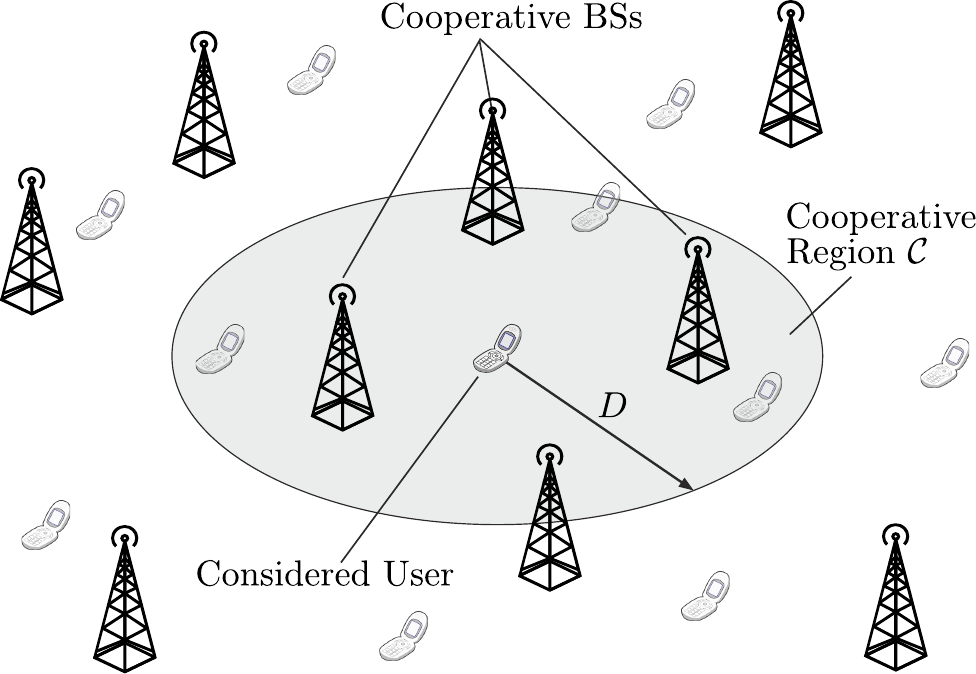}
\caption{Illustration of the considered scenario. A cluster of cooperative BSs is formed around the typical user.}
\label{fig:illustration}
\end{figure}

\section{Cooperation Scenario and System Model}\label{sec:model}
We consider a single-tier OFDMA-based cellular system in the downlink with single-antenna BSs scattered in the plane according to a stationary Poisson point process \cite{stoyan95} (PPP) $\Phi\mathdef\{\mathsf{x}_{i}\}_{i=1}^{\infty}$ with density $\lambda$, where $\mathsf{x}_{i}\in\mathbb{R}^2$ denotes the random location of the $i$-th BS. While the PPP assumption inherently neglects the correlation in the BS locations present in real cellular deployments, it offers analytical tractability, and has therefore become well-accepted for modeling cellular networks \cite{andrews11,blas12}. We assume single-antenna users (receivers) distributed according to a PPP and focus the analysis on a typical user located at the origin $o\in\mathbb{R}^2$. By Slivnyak's Theorem\cite{stoyan95} and due to the stationarity of $\Phi$, the typical user will reflect the spatially-averaged downlink performance under the metrics defined in Section~\ref{sec:metrics}. To each BS, we further assign a mark $\mathsf{g}_{i}\in\mathbb{R}^{+}$ denoting the power fading on the channel from the $i$-th BS to the typical user. We assume independent and identically distributed (i.i.d.) narrow-band fading. The marks follow an identical law with probability density function (PDF) $f_{\mathsf{g}}(g)$, where we assume $\mathbb{E}\left[\mathsf{g}\right]=1$ and $\mathbb{E}\left[\mathsf{g}^2\right]<\infty$. The considered scenario is shown in Fig.~\ref{fig:illustration}.

\subsection{Cooperation and Cluster Model}
We consider a \textit{user-centric} BS clustering scheme, in which a cooperative set of BSs is assigned to each user. Whether a BS is included in the cooperative set of a given user is usually based on long-term (fading-averaged) received signal strength (RSS) measurements reported on the uplink. Hence, for the typical user located at the origin, BSs that are sufficiently close are grouped into a cooperative cluster. Making this mathematically strict: BSs inside the \emph{cooperative region} $\mathcal{C}\subseteq\mathbb{R}^2$, defined by the two-dimensional ball $\mathcal{C}=b(o,D)$, are members of the cooperative cluster of the typical user. Thereby, the radius $D$ controls the size of $\mathcal{C}$, and hence serves as a tunable parameter for balancing enhanced user experience through more cooperation on the one hand and increased backhaul complexity due to more overhead on the other hand.

In NC-JT, a user receives a non-coherent sum of multiple copies of the useful signal transmitted by the cooperating BSs. BSs of the same cooperative cluster not actively participating in a joint transmission may still cooperate indirectly by not transmitting on the same radio resources used for joint transmission. This indirect form of cooperation can be seen as intra-cluster coordinated scheduling (CS). Alternatively, cooperative BSs not participating in an ongoing joint transmission may reuse the radio resources allocated to the joint transmission, which, in turn, creates intra-cluster interference to the NC-JT user. This form of intra-cluster scheduling can be seen as intra-cluster frequency reuse (FR). For better exposition of the results, we assume intra-cluster FR throughout this work except for Section~\ref{sec:scheduling}, where the two scheduling schemes are compared. At the receiver, the non-coherent sum of useful signals leads to cyclic delay diversity, which translates into a received power boost\cite{docomo10,morimoto06,gosh10}. In order to obtain this increase, the receive filter must be matched to the \emph{composite} channel, hence requiring CSI at the receiver (CSI-R). The effect of imperfect CSI-R is treated in Section \ref{sec:csi}. See Appendix~\ref{sec:nc_jt_scheme} for more details about the transmission/reception procedure in NC-JT.

\subsection{Channel-Dependent Cooperation Activation}
CSI must be gathered by the cooperative BSs to decide whether a user should be served jointly. Among the different components determining the outcome of this decision process, the individual \emph{instantaneous} RSSs to all BSs of the cooperative cluster have major influence; clearly, if the channel to a cooperative BS is in a deep fade, information cannot be sent reliably over that link. In this case, a cooperative BS will defer from assisting the joint transmission.

To capture this influential mechanism, we hence assume that a cooperative BS gets engaged in a joint transmission only if the instantaneous RSS on the corresponding link is above a threshold $T\geq0$. In some cases, it will be useful to express $T$ with respect to the received power from a hypothetical BS located at the cluster edge, i.e., $T=\tilde T D^{-\alpha}$ with $\tilde T$ being the cluster-edge representation of $T$. Other aspects related to the activation of cooperation include for instance the optimization metric and the cost function for quantifying the gains of setting-up a cooperative transmission, cf. \cite{marsch11} for further discussions. Capturing all these aspects while preserving analytical tractability is outside the scope of this work. Table~\ref{tab:notation} summarizes the notation used in this work.

\subsection{Performance Metrics: \texorpdfstring{$\mathtt{SINR}$}{SINR} and Spectral Efficiency}\label{sec:metrics}
The main purpose of downlink BS cooperation is to increase throughput for users experiencing a hostile radio environment \cite{irmer11}, i.e., cell-edge users with low $\mathtt{SINR}$; hence the $\mathtt{SINR}$ is an important metric. In highly loaded cells, the rate offered to a user and the $\mathtt{SINR}$ are tightly coupled via the cell load \cite{heath12,singh13}, which is difficult to model and analyze in a cooperative scenario due to the fact that scheduling decisions are made across different cells. In contrast, the rate offered to a user in a lightly loaded cell is only limited by the maximum channel bandwidth of the user frontend, e.g., $20$~MHz in LTE \cite{gosh10}. In such a lightly-loaded cell scenario the effect of used density, and hence cell load, is of subordinate importance and the spectral efficiency $\mathtt{R}$ is an appropriate performance metric. For NC-JT in particular, the range of practical scenarios is limited to lightly loaded cells only \cite{ericsson11}. In this work, we will hence assume a lightly-loaded cell scenario and use the $\mathtt{SINR}$ and the spectral efficiency $\mathtt{R}$ as performance metrics. Note that in the high-load scenario, these metrics cannot solely reflect the overall performance as the cell load and the limitation of resources are not captured.  

Under the NC-JT Tx/Rx scheme explained in Appendix~\ref{sec:nc_jt_scheme}, the $\mathtt{SINR}$ at the typical user can be expressed as\footnote{Note that \eqref{eq:sinr_def} does not correspond to the actual $\mathtt{SINR}$ on one resource element but to the average $\mathtt{SINR}$ experienced on coherent subcarriers taking into account the time/phase mismatch of NC-JT, cf. Appendix~\ref{sec:nc_jt_scheme}.}
\begin{IEEEeqnarray}{rCl}
	\mathtt{SINR} &\mathdef& \frac{\sum\limits_{i\in\Phi\cap\mathcal{C}}\mathsf{g}_{i}\|\mathsf{x}_{i}\|^{-\alpha}\mathds{1}(\mathsf{g}_{i}\|\mathsf{x}_{i}\|^{-\alpha}\geq T)}
	{\hspace{-.1cm}\sum\limits_{i\in\Phi\cap\mathcal{C}}\hspace{-.15cm} \mathsf{g}_{i}\|\mathsf{x}_{i}\|^{-\alpha}\mathds{1}(\mathsf{g}_{i}\|\mathsf{x}_{i}\|^{-\alpha}\hspace{-.1cm}< T)+\hspace{-.2cm} \sum\limits_{i\in\Phi\cap\mathcal{\bar C}}\hspace{-.15cm}\mathsf{g}_{i}\|\mathsf{x}_{i}\|^{-\alpha}+\frac{1}{\eta}}\IEEEnonumber\\
	&=&\frac{\sig}{\mathsf{J}_{\mathcal{C}}+\mathsf{J}_{\mathcal{\bar C}}+\frac{1}{\eta}},\label{eq:sinr_nc}\IEEEeqnarraynumspace\label{eq:sinr_def}
\end{IEEEeqnarray}
where 
\begin{itemize}
	\item $\|x_{i}\|^{-\alpha}$: path loss to the $i$-th BS; $\alpha>2$ is the path loss exponent.
	\item $\eta$: transmit signal-to-noise ratio (BS transmit power divided by receiver noise). Receiver noise is modeled as additive white Gaussian noise (AWGN).
	\item $\sig\mathdef\sum_{i\in\Phi\cap\mathcal{C}}\mathsf{g}_{i}\|\mathsf{x}_{i}\|^{-\alpha}\mathds{1}(\mathsf{g}_{i}\|\mathsf{x}_{i}\|^{-\alpha}\geq T)$: received signal power from the cooperating BSs in $\mathcal{C}$ that \emph{serve the typical user}.
	\item $\mathsf{J}_{\mathcal{C}}\mathdef\sum_{i\in\Phi\cap\mathcal{C}} \mathsf{g}_{i}\|\mathsf{x}_{i}\|^{-\alpha}\mathds{1}(\mathsf{g}_{i}\|\mathsf{x}_{i}\|^{-\alpha}< T)$: sum interference (power) caused by cooperating BS in $\mathcal{C}$ that \emph{serve other users} on the same resource (for intra-cluster FR); $\mathsf{J}_{\mathcal{C}}\equiv0$ for intra-cluster CS and/or $T=0$.
	\item $\mathsf{J}_{\mathcal{\bar C}}\mathdef\sum_{i\in\Phi\cap\mathcal{\bar C}}\mathsf{g}_{i}\|\mathsf{x}_{i}\|^{-\alpha}$: sum interference (power) created by BSs outside $\mathcal{C}$ ($\mathcal{\bar C}=\mathbb{R}^2\setminus \mathcal{C}$).
\end{itemize}

Treating interference as noise and assuming capacity-achieving codes, the spectral efficiency $\mathtt{R}$ at the typical user is given by $\mathtt{R}\mathdef\log_{2}(1+\mathtt{SINR})$ in bit/s/Hz. Note that for analytical tractability we shall assume that all out-of-cluster BSs create interference to the considered user. Since we consider a lightly-loaded cell scenario, this assumption may tendentially overestimate the actual interference.

\begin{table}[!t]
	\renewcommand{\arraystretch}{1.3}
	\caption{Notation used in this work}
	\label{tab:notation}
	\centering
	\small
	\begin{tabular}{c|p{6.5cm}}
		\hline
		\bfseries{Notation} & \hspace{2.4cm}\bfseries{Description}\\
		\hline
		$\Phi;\lambda$ 		& BS location process $\Phi$ with average density $\lambda$\\
		\hline
		$\alpha$			& Path loss exponent\\
		\hline
		$\mathcal{C};D$	& Cooperative region $\mathcal{C}=b(o,D)$ with radius $D$\\
		\hline
		$K$				& Average number of BSs in $\mathcal{C}$, i.e., cooperative BSs\\
		\hline
		$T;\tilde T$			& Cooperation activation threshold $T$; $\tilde T=D^{\alpha}T$\\
		\hline
		$\sig$				& Useful received signal power at the typical user\\
		\hline
		\begin{minipage}[t][1em][c]{1cm}$\mathsf{J}_{\mathcal{C}};\mathsf{J}_{\mathcal{\bar C}};\\\mathsf{J}_{\text{CSI}}$\end{minipage} & Intra-cluster interference; out-of-cluster interference; residual interference due to imperfect CSI\\
		\hline
		\begin{minipage}[t][1em][c]{1cm}$\mathtt{SINR}$;\\$\mathtt{SINR}_{\text{pilot},i}$\end{minipage}& $\mathtt{SINR}$ under NC-JT; Channel-estimation $\mathtt{SINR}$ of the $i$-th cooperative-BS link to the typical user\\
		\hline
		$\mathtt{R}$		& Spectral efficiency $\log_{2}(1+\mathtt{SINR})$\\
		\hline
		$\eta$				& Downlink transmit $\mathtt{SNR}$ on one resource element\\
		\hline
		$k,\theta$			& Shape $k$ and scale $\theta$ of Gamma distribution\\
		\hline
		$\sigma^{2}_{\text{MMSE},i}$ & Receiver-side channel-estimation MMSE of the link to the $i$-th cooperating BS\\
		\hline
		$N_{\text{pilot}}$ & Total number of sampling points (pilots) for channel estimation during one channel coherence period\\
		\hline
		$\Delta$			& Average radio resource saving when switching from CS to FR intra-cluster scheduling\\
		\hline
	\end{tabular}
\end{table}

\section{\texorpdfstring{$\mathtt{SINR}$}{SINR} Characterization}\label{sec:characterization}
Note that in \eqref{eq:sinr_nc}, $\sig$ and $\mathsf{J}_{\mathcal{C}}$ are statistically independent due to the mutually disjoint events inside the indicator functions \cite{last95}. Since $\Phi$ has Poisson property, $\mathsf{J}_{\mathcal{\bar C}}$ is statistically independent from $\sig$ and $\mathsf{J}_{\mathcal{C}}$. Still, the expression in \eqref{eq:sinr_nc} is difficult to work with directly. To get a better handle on the $\mathtt{SINR}$, we will next approximate the compound term in the denominator.

\subsection{Interference-plus-Noise Gamma Approximation}
In \cite{ganti09}, the authors showed that the Gamma distribution can provide a reasonably tight fit to the statistics of Poisson interference. A Gamma approximation of the interference was also used in \cite{heath12}. Motivated by these findings, we approximate the denominator in \eqref{eq:sinr_nc} by a Gamma random variable, whose shape $k$ and scale $\theta$ can be obtained through second-order moment matching.
\begin{proposition}\label{prop:interference_ap}
	The denominator in \eqref{eq:sinr_nc} can be approximated by a Gamma random variable $\mathsf{\tilde J}\approx\mathsf{J}_{\mathcal{C}}+\mathsf{J}_{\mathcal{\bar C}}+\frac{1}{\eta}$ with distribution $\mathbb{P}(\mathsf{\tilde{J}}\leq z)=1-\Gamma(k,z/\theta)/\Gamma(k)$, where $k$ and $\theta$ are
	\begin{IEEEeqnarray}{rCl}
		k = 4\pi\lambda\frac{\alpha-1}{(\alpha-2)^2}\frac{\left(\mathbb{E}\left[\mathsf{g}\min\{D^{\alpha},\frac{\mathsf{g}}{T}\}^{\frac{2}{\alpha}-1}\right]+\frac{\alpha-2}{2\pi\lambda\eta}\right)^2}{\mathbb{E}\left[\mathsf{g}^2\min\{D^{\alpha},\tfrac{\mathsf{g}}{T}\}^{\frac{2}{\alpha}-2}\right]}\IEEEeqnarraynumspace\label{eq:shape}
	\end{IEEEeqnarray}
	and
	\begin{IEEEeqnarray}{rCl}
	\theta&=&\frac{1}{2}\frac{\alpha-2}{\alpha-1}\frac{\mathbb{E}\left[\mathsf{g}^2\min\{D^{\alpha},\tfrac{\mathsf{g}}{T}\}^{\frac{2}{\alpha}-2}\right]}{\mathbb{E}\left[\mathsf{g}\min\{D^{\alpha},\tfrac{\mathsf{g}}{T}\}^{\frac{2}{\alpha}-1}\right]+\frac{\alpha-2}{2\pi\lambda\eta}},\IEEEeqnarraynumspace\label{eq:scale}
	\end{IEEEeqnarray}
	where $\Gamma(a,x)=\int_{x}^{\infty}t^{a-1}e^{-xt}\,\mathrm dx$ is the upper incomplete Gamma function. 
\end{proposition}
\begin{IEEEproof}
See Appendix~\ref{ap:interference_ap}.
\end{IEEEproof}
For the following special cases, \eqref{eq:shape} and \eqref{eq:scale} can be further simplified by explicitly computing
\begin{IEEEeqnarray}{rCl}
&&\mathbb{E}\left[\mathsf{g}\min\left\{D^{\alpha},\frac{\mathsf{g}}{T}\right\}^{\frac{2}{\alpha}-1}\right]\IEEEnonumber\\
&&\quad\quad\quad\quad=\begin{cases}
		T^{1-\frac{2}{\alpha}}\gamma(1+\tfrac{2}{\alpha},TD^{\alpha})\\\quad+D^{2-\alpha}\Gamma(2,TD^{\alpha}), & \mathsf{g}\sim\text{Exp}(1),\\
				D^{2-\alpha}, & T=0,
		\end{cases}\IEEEeqnarraynumspace\label{eq:first_moment_simple}
\end{IEEEeqnarray}
where $\gamma(a,x)=\Gamma(a)-\Gamma(a,x)$ is the lower incomplete Gamma function, and
\begin{IEEEeqnarray}{rCl}
&&\mathbb{E}\left[\mathsf{g}^2\min\left\{D^{\alpha},\frac{\mathsf{g}}{T}\right\}^{\frac{2}{\alpha}-2}\right]\IEEEnonumber\\
&&\quad\quad\quad\quad=\begin{cases}
		T^{2-\frac{2}{\alpha}}\gamma(1+\tfrac{2}{\alpha},TD^{\alpha})\\\quad+D^{2-2\alpha}\Gamma(3,TD^{\alpha}), & \mathsf{g}\sim\text{Exp}(1),\\
				\mathbb{E}\left[\mathsf{g}^2\right]D^{2-2\alpha}, & T=0.
		\end{cases}\label{eq:variance_simple}
\end{IEEEeqnarray}
\begin{figure*}[!t]
	\centerline{\subfloat[$\alpha=3$]{\includegraphics[width=0.487\textwidth]{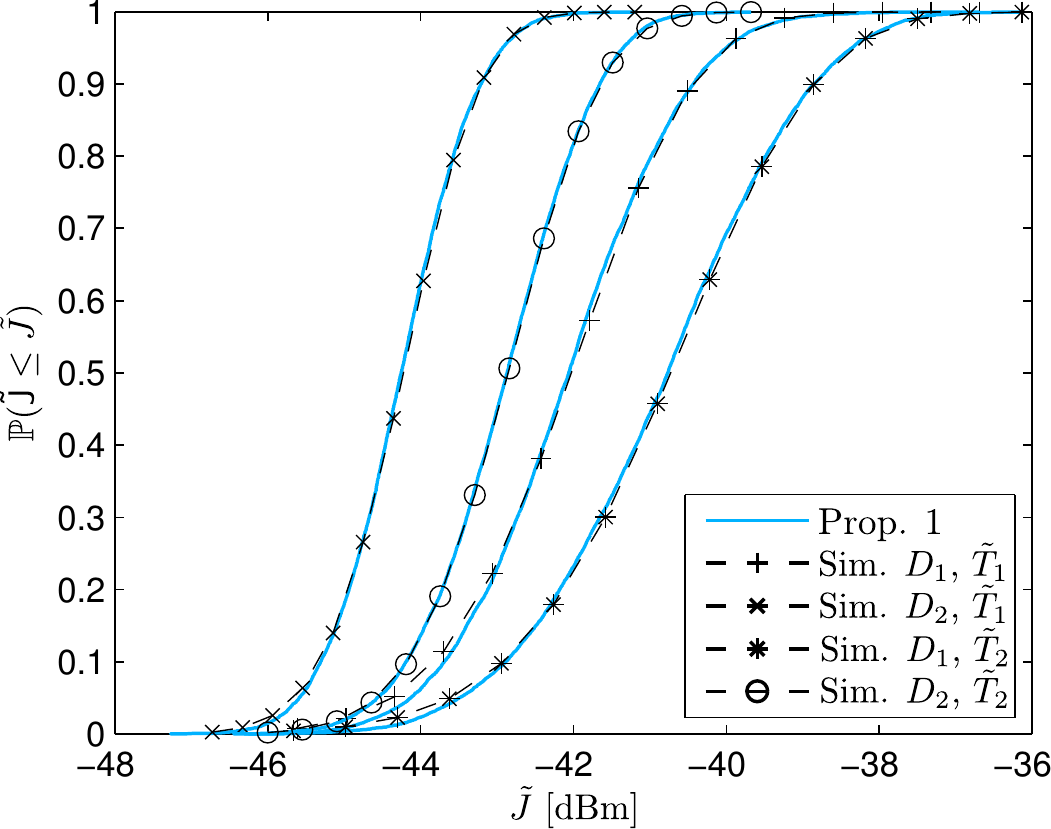}
	\label{fig:inter_approx1}}
	\hfil
	\subfloat[$\alpha=5$]{\includegraphics[width=0.487\textwidth]{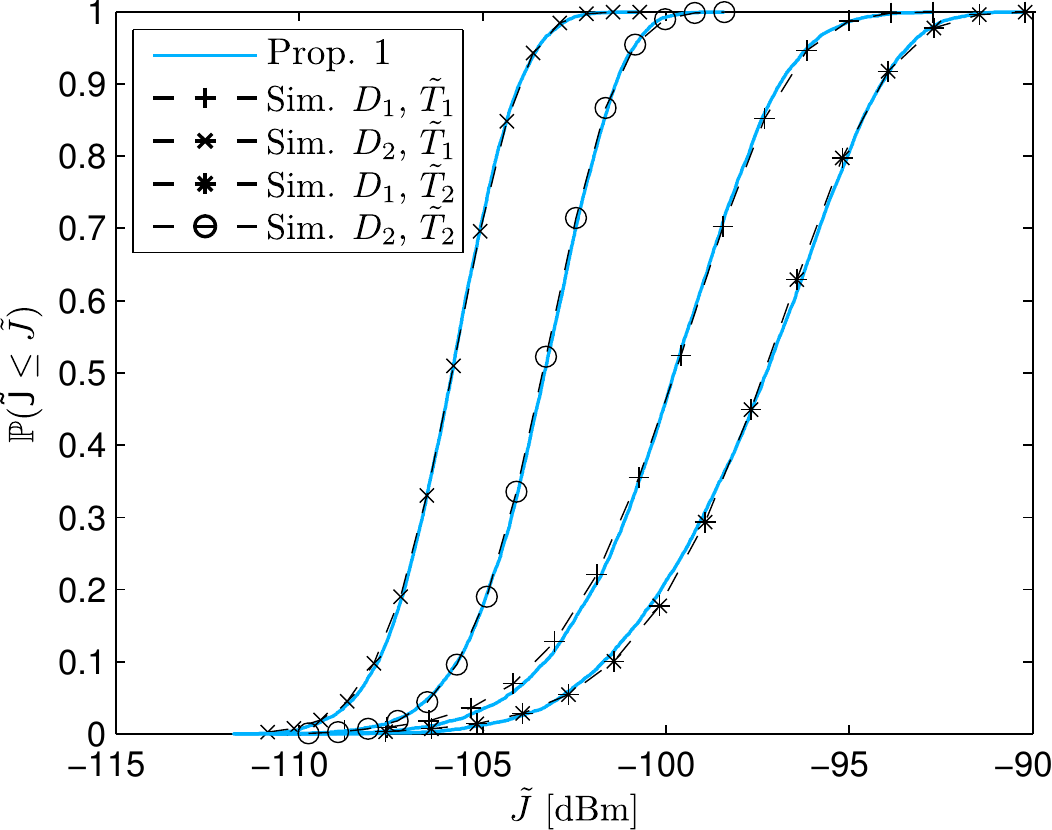}
	\label{fig:inter_approx2}}}
	\caption{Estimated CDF (dashed) and Gamma CDF (solid with marks) for $\alpha=3$ (a) and $\alpha=5$ (b). Average inter-BS distance is $\sim500$\,m. Parameters are as follows:  $D_{1}=450$\,m ($\sim3$ cooperative BSs on average) and $D_{2}=750$\,m ($\sim8$ cooperative BSs on average), $\eta=162$\,dB. $\tilde T_{1}=0$\,dB and $\tilde T_{2}=6$\,dB. Due to variation of $D$ and $\alpha$, absolute value $T$ changes accordingly.}
\end{figure*}
Observe that \eqref{eq:first_moment_simple} and \eqref{eq:variance_simple} tend to zero as $D\to\infty$, i.e., increasing the cooperative region $\mathcal{C}$ converts more interfering BSs into cooperative BSs. For $T=0$ interference is caused only by BSs outside the cooperative cluster ($\mathsf{J}_{\mathcal{C}}=0$). This corresponds to the case of all BSs in $\mathcal{C}$ always serving the users jointly irrespective of the current channel realizations. When letting $D\to0$, it is seen that both moments do not exist, which is due to the singularity of the path loss law \cite{ganti09}.

Fig. \ref{fig:inter_approx1} and Fig. \ref{fig:inter_approx2} compare the cumulative distribution function (CDF) given in Proposition~\ref{prop:interference_ap} with the empirical CDF obtained through simulations. The transmit $\mathtt{SNR}$ was set to $\eta=162$\,dB, which is a typical value in LTE networks \cite{holma07}. As can be seen, the Gamma approximation provides a good fit to the true CDF for a wide range of system parameters.
\begin{remark}
Note that the approximation becomes obsolete when $\frac{1}{\eta}\gg\mathsf{J}_{\mathcal{C}}+\mathsf{J}_{\mathcal{\bar C}}$, since the denominator in \eqref{eq:sinr_nc} degenerates to $\frac{1}{\eta}$ (noise-limited case). As interference is usually the performance-limiting factor in today's cellular networks, we will avoid this pathological case in our analysis.
\end{remark}
\begin{remark}\label{rem:fading}
For certain fading statistics, e.g., deterministic or Nakagami-Lognormal, the Gamma approximation may not preserve the interference tail behavior, cf.\cite[Sec.~5.4.2]{ganti09}. The resulting approximation error is studied in Section~\ref{sec:main_result}.
\end{remark}
\subsection{Main Result: \texorpdfstring{$\mathtt{SINR}$}{SINR} Distribution}\label{sec:main_result}
\begin{theorem}\label{thm:sinr_distribution}
The $\mathtt{SINR}$ distribution for the typical user at the origin, in the described setting, is bounded above and below as
\begin{IEEEeqnarray}{rCl}
	\mathbb{P}(\mathtt{SINR}\leq\beta)\overset{\tilde k=\lceil k\rceil}{\underset{\tilde k= \lfloor k \rfloor}\lesseqgtr} \sum\limits_{m=0}^{\tilde k-1}\lap_{\sig}^{(m)}(\tfrac{1}{\theta\beta})\frac{(\theta\beta)^{-m}}{m!},
	\IEEEeqnarraynumspace\label{eq:sinr_distribution}
\end{IEEEeqnarray}
where $\mathcal{L}_{\sig}(s)\mathdef\mathbb{E}\left[e^{-s\sig}\right]$ is the Laplace transform of $\sig$, $\lap_\sig^{(m)}(s_{0})\mathdef\partial^{m} \lap_\sig(-s)/\partial s^m|_{s=-s_{0}}$ is the $m$-th derivative of $\lap_{\sig}(-s)$ evaluated at $s=-s_{0}$.
\end{theorem}
\begin{IEEEproof}
See Appendix~\ref{ap:sinr_distribution}.
\end{IEEEproof}
Before specializing the result of Theorem~\ref{thm:sinr_distribution} to certain concrete cases, we will discuss some properties of \eqref{eq:sinr_distribution} below.

\textit{Generality:} In many cases, the PDF of $\sig$ is not known while its Laplace transform can be given in closed-form. As \eqref{eq:sinr_distribution} requires only the Laplace transform of $\sig$, more specifically its $\tilde k-1$ first derivatives, the generality offered by this result is evident. Moreover, we do not have to specify the fading PDF $f_{\mathsf{g}}$. Furthermore, the superposition property of Laplace transforms, i.e., $\mathcal{L}_{\sum_{i}f_{i}}=\prod_{i}\mathcal{L}_{f,i}$ for i.i.d. $f_{i}$, can be readily exploited by the convenient form of Theorem~\ref{thm:sinr_distribution}.

\textit{Tightness of bounds:} The reason why \eqref{eq:sinr_distribution} is given in terms of upper/lower bounds is due to the necessity of rounding $k$ to an integer; the sum in \eqref{eq:sinr_distribution} is truncated at $\lfloor k\rfloor$ (lower bound) and extended to $\lceil k \rceil$ (upper bound). A straightforward way to study the tightness of the bounds is to characterize the gap between the lower and upper bound. The worst-case gap is equal to last summand in \eqref{eq:sinr_distribution}. Whenever $k$ is integer-valued, either the upper or the lower bound becomes exact.

\textit{Effect of path loss:} Intuitively, the interference created by the many far BSs is more harmful for small $\alpha$. At the same time, however, the useful signals undergo a milder path loss. For non-cooperative transmission, it is known that as $\alpha\to2$, the $\mathtt{SINR}$ tends to zero a.s. \cite{ganti09}. Observe from \eqref{eq:shape} and \eqref{eq:scale} that $\alpha\to2$ implies $k\to\infty$ and $\theta\to0$. Taking the limit $\alpha\to2$ in \eqref{eq:sinr_distribution}, we have
\begin{IEEEeqnarray}{rCl}
	 &&\lim\limits_{\alpha\to2}\sum\limits_{m=0}^{\tilde k-1}\lap_{\sig}^{(m)}(\tfrac{1}{\theta\beta})\frac{(\theta\beta)^{-m}}{m!}\IEEEnonumber\\
	 &&\quad\quad\overset{\text{(a)}}{=} \lim\limits_{\alpha\to2} \sum\limits_{m=0}^{\tilde k-1}\frac{(\theta\beta)^{-m}}{m!}\int_{0}^{\infty}P^{m}\,f_{\sig}(P)\,e^{-\frac{P}{\theta\beta}}\,\mathrm dP\IEEEnonumber\\
	&&\quad\quad\overset{\text{(b)}}{=}\lim\limits_{\alpha\to2}\theta\beta \sum\limits_{m=0}^{\tilde k-1}\frac{1}{m!}\int_{0}^{\infty}u^{m}\,f_{\sig}(u\theta\beta)\,e^{-u}\,\mathrm du\IEEEnonumber\\
	&&\quad\quad\overset{\text{(c)}}{=} \lim\limits_{\alpha\to2}\theta\beta\int_{0}^{\infty}f_{\sig}(u\theta\beta)\,e^{-u}\sum\limits_{m=0}^{\tilde k-1}\frac{u^{m}}{m!}\,\mathrm du\IEEEnonumber\\
	&&\quad\quad\overset{\text{(d)}}{=} 1-\lim\limits_{\alpha\to2}\int_{0}^{\infty}f_{\sig}(P)\,\frac{\gamma(\tilde{k}+1,P/\theta\beta)}{\Gamma(\tilde{k}+1)}\,\mathrm dP\IEEEnonumber\\
	&&\quad\quad\overset{\text{(e)}}{=} 1-\int_{0}^{\infty}\underbrace{\lim\limits_{\alpha\to2}f_{\sig}(P)\frac{\gamma(\tilde{k}+1,P/\theta\beta)}{\Gamma(\tilde{k}+1)}}_{\to 0}\,\mathrm dP=1.\IEEEeqnarraynumspace\label{eq:freq_ref}
\end{IEEEeqnarray}
(a) follows from the $s$-differentiation theorem for the Laplace transform\cite{lepage80}, (b)  follows from the substitution $P/\theta\beta\to u$, (c) follows from Tonelli's theorem \cite{bauer92}, (d) follows from the substitution $u\theta\beta\to P$, and (e) follows from the dominated convergence theorem ($0\leq\gamma(\tilde{k}+1,P/\theta\beta)/\Gamma(\tilde{k}+1)\leq1$ for all $P,\alpha$) and from the fact that $\gamma(a,z)/\Gamma(a)\sim (2\pi a)^{-\frac{1}{2}}e^{a-z}(z/a)^{a}\to0$ as $a\to\infty$ \cite{olver10}. Thus, $\mathbb{P}(\mathtt{SINR}\leq\beta)\to1$ as $\alpha\to2$, thereby showing that the interference created by the many far BSs indeed outweighs the milder path loss of the cooperative links. Conversely, we expect the $\mathtt{SINR}$ to improve with larger $\alpha$ in accordance with the literature\cite{andrews11}.

Using a linear combination of the upper and lower bound in \eqref{eq:sinr_distribution} with weights chosen according to the relative distance of $k$ to $\lfloor k \rfloor$ and $\lceil k \rceil$, we get the following approximation for $\mathbb{P}(\mathtt{SINR}\leq\beta)$.
\begin{corollary}\label{col:approx_sinr}
The $\mathtt{SINR}$ distributed can be approximated by
	\begin{IEEEeqnarray}{rCl}
		\mathbb{P}(\mathtt{SINR}\leq\beta)&\approx& (k-\lfloor k \rfloor)\lap_{\sig}^{(\lceil k\rceil)}(\tfrac{1}{\theta\beta})\frac{(\theta\beta)^{-\lceil k \rceil}}{\lceil k \rceil!}\IEEEnonumber\\
		&&+\sum\limits_{m=0}^{\lfloor k \rfloor}\lap_{\sig}^{(m)}(\tfrac{1}{\theta\beta})\frac{(\theta\beta)^{-m}}{m!}.\IEEEeqnarraynumspace\label{eq:cdf_ap}
	\end{IEEEeqnarray}
\end{corollary}
Although the approximation in \eqref{eq:cdf_ap} looks rather simple, it turns out that it is remarkably tight as will be demonstrated later. We propose a second alternative to Theorem~\ref{thm:sinr_distribution}, which is useful when the $\tilde k,\tilde k+1,\ldots$-th derivatives of $\lap_\sig$, can be easily estimated or bounded.
\begin{corollary}
Let $\zeta(\tilde k)$ be an (arbitrarily good) estimate of the sum $\sum_{m=\tilde k}^{\infty}\mathcal{L}_{\sig}^{(m)}(\tfrac{1}{\theta\beta})\frac{(\theta\beta)^{-m}}{m!}$. Then, one has $\mathbb{P}(\mathtt{SINR}\leq\beta)\approx1-\zeta(\tilde k)$, which yields an (arbitrarily good) approximation.
\end{corollary}
\begin{IEEEproof}
	We modify \eqref{eq:sinr_distribution} as follows:
	\begin{IEEEeqnarray}{rCl}
		\mathbb{P}(\mathtt{SINR}\leq\beta)&=& \sum\limits_{m=0}^{\infty}\lap_{\sig}^{(m)}(\tfrac{1}{\theta\beta})\frac{(\theta\beta)^{-m}}{m!}\IEEEnonumber\\
		&&-\sum\limits_{m=\tilde k}^{\infty}\lap_{\sig}^{(m)}(\tfrac{1}{\theta\beta})\frac{(\theta\beta)^{-m}}{m!}\approx1-\zeta(\tilde k),\IEEEeqnarraynumspace
	\end{IEEEeqnarray}
	where the first sum is equal to one as a result of Hille's theorem \cite{feller71}.
\end{IEEEproof}

To discuss further properties of \eqref{eq:sinr_distribution}, we next specify the form of $\lap_\sig$ for two cases of interest.

\subsubsection{Fixed Number of Cooperating BSs}\label{sec:bpp}
We assume that the number of cooperating BSs in $\mathcal{C}$ is equal to $K>0$ which is equivalent to conditioning the PPP on $\Phi(\mathcal{C})=K$.\footnote{With a small abuse of notation, we define $\Phi(\mathcal{C})\mathdef \sum_{i\in\Phi}\mathds{1}(\mathsf{x}_{i}\in\mathcal{C})$ as the random counting measure.} Fixing the number of cooperating BSs will offer the possibility to relate our $\mathtt{SINR}$ results to other works that commonly assume a certain cluster size $K$. In this case, the locations of the $K$ BSs then follows a Binomial point process (BPP) \cite{stoyan95} on $\mathcal{C}$. We refer to this case as the \emph{conditional case}. Note that conditioned on $\Phi(\mathcal{C})=K$, $\sig$ and $\mathsf{J}_{\mathcal{C}}$ are now negatively correlated. We will however treat them as being statistically independent and interpret the resulting error as additional inaccuracy of the Gamma approximation.

\begin{lemma}\label{lem:laplace_sig_bpp}
The Laplace transform $\lap_{\sig|\Phi(\mathcal{C})=K}(s)$ (conditional case) is given by
\begin{IEEEeqnarray}{rCl}
	&&\lap_{\sig|\Phi(\mathcal{C})=K}(s)\IEEEnonumber\\
	&&\quad= \Big(1-\frac{1}{D^2}\mathbb{E}\left[\min\{D^{\alpha},\tfrac{\mathsf{g}}{T}\}^{\frac{2}{\alpha}}\left(1-e^{-s\mathsf{g}\max\{D^{-\alpha},\frac{T}{\mathsf{g}}\}}\right)\right.\IEEEnonumber\\
	&&\quad\quad\quad\quad\quad\left.+(s\mathsf{g})^{\frac{2}{\alpha}}\Gamma(1-\tfrac{2}{\alpha},s\mathsf{g}\max\{D^{-\alpha},\tfrac{T}{\mathsf{g}}\})\right]\Big)^K.\IEEEeqnarraynumspace\label{eq:lap_bpp}
	\end{IEEEeqnarray}
\end{lemma}
\begin{IEEEproof}
See Appendix~\ref{ap:laplace_sig_bpp}.
\end{IEEEproof}

\subsubsection{Poisson Number of Cooperating BSs}\label{sec:ppp}
The number of cooperative BSs inside $\mathcal{C}$ is now assumed random and given by $\Phi(\mathcal{C})$ following a Poisson distribution. We refer to this case as the \emph{unconditional case}. By deconditioning $\lap_{\sig|\Phi(\mathcal{C})=K}(s)$ on $K$, we obtain the equivalent result for the unconditional case.

\begin{lemma}\label{lem:laplace_sig_ppp}
The Laplace transform $\lap_{\sig}(s)$ (unconditional case) is given by
\begin{IEEEeqnarray}{rCl}
\lap_{\sig}(s)&=&\exp\hspace{-.05cm}\Big\{\hspace{-.07cm}-\hspace{-.05cm}\lambda\pi\mathbb{E}\hspace{-.05cm}\left[\min\{D^{\alpha}\hspace{-.02cm},\tfrac{\mathsf{g}}{T}\}^{\frac{2}{\alpha}}\hspace{-.06cm}\left(1-e^{-s\mathsf{g}\max\{D^{-\alpha}\hspace{-.07cm},\frac{T}{\mathsf{g}}\}}\hspace{-.02cm}\right)\right.\IEEEnonumber\\
&&\qquad\left.+(s\mathsf{g})^{\frac{2}{\alpha}}\Gamma(1-\tfrac{2}{\alpha},s\mathsf{g}\max\{D^{-\alpha},\tfrac{T}{\mathsf{g}}\})\right]\hspace{-.05cm}\Big\}.\IEEEeqnarraynumspace\label{eq:lap_ppp}
\end{IEEEeqnarray}
\end{lemma}
\begin{IEEEproof}
See Appendix~\ref{ap:laplace_sig_ppp}.
\end{IEEEproof}

\textit{Effect of BS density:} Increasing the BS density $\lambda$ has two opposing effects: 1) it causes more interference, since the number of active BSs in the network is increased; 2) it increases the chances of being jointly served by multiple cooperative BSs which manifests itself in decay of $\lap_\sig$. To understand the underlying trend, we study the behavior of \eqref{eq:sinr_distribution} as $\lambda\to\infty$. In this limit it suffices to treat the unconditional case since \eqref{eq:lap_bpp} and \eqref{eq:lap_ppp} then become equal\cite{stoyan95}.

Noting from \eqref{eq:lap_ppp} that $\lap_{\sig}$ is of the form $e^{-\lambda f(s)}$, where $f(s)$ does not depend on $\lambda$, the corresponding $m$-th derivative with respect to $s$ must be of the form $e^{-\lambda f(s)}h_{m}(\lambda,s)$, where $h_{m}(\lambda,s)$ is polynomial in $\lambda$. Hence, we can rewrite \eqref{eq:sinr_distribution} in the form $e^{-\lambda f(1/\theta\beta)}\sum_{m=0}^{\tilde k-1}\hspace{-.1cm}\frac{(\theta\beta)^{-m}}{m!}h_{m}(\lambda,1/\theta\beta).$ Now observe that the leading term $e^{-\lambda f(1/\theta\beta)}$ is exponentially decreasing in $\lambda$, and hence dominates the scaling of \eqref{eq:sinr_distribution} as $\lambda\to\infty$ for $\theta,\beta>0$. This means that the $\mathtt{SINR}$ can be increased by adding more BSs. Remarkably, this $\mathtt{SINR}$ gain is achieved without the need for careful deployment since increasing $\lambda$ means adding both more cooperative \emph{and} more interfering BSs. This finding is somewhat interesting in spite of recent results \cite{dhillon12}, showing that in a single-tier cellular network with $\max$-power association and no cooperation, the BS density does not affect the $\mathtt{SIR}$ distribution. In the cooperative scenario, in contrast, a denser deployment of BSs may be beneficial. A similar observation was made in \cite{huang11} for CB/CS BS cooperation. This finding, however, must be treated with care as it assumes a fixed $D$, implying the cluster size $K$ to increase as well.

\begin{figure*}[!t]
	\centerline{\subfloat[$\mathtt{SINR}$ CDF (unconditional case)]{\includegraphics[width=0.495\textwidth]
	{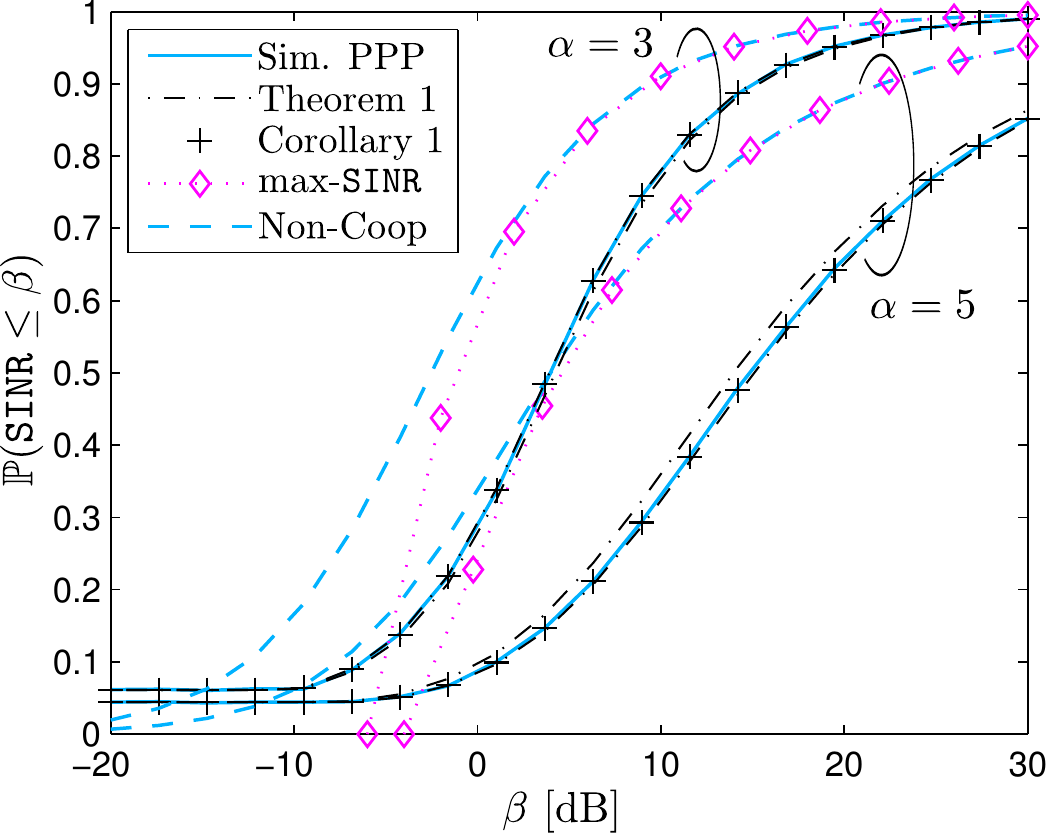}
	\label{fig:inter_sir1}}
	\hfil
	\subfloat[$\mathtt{SINR}$ CDF (conditional case)]{\includegraphics[width=0.495\textwidth]
	{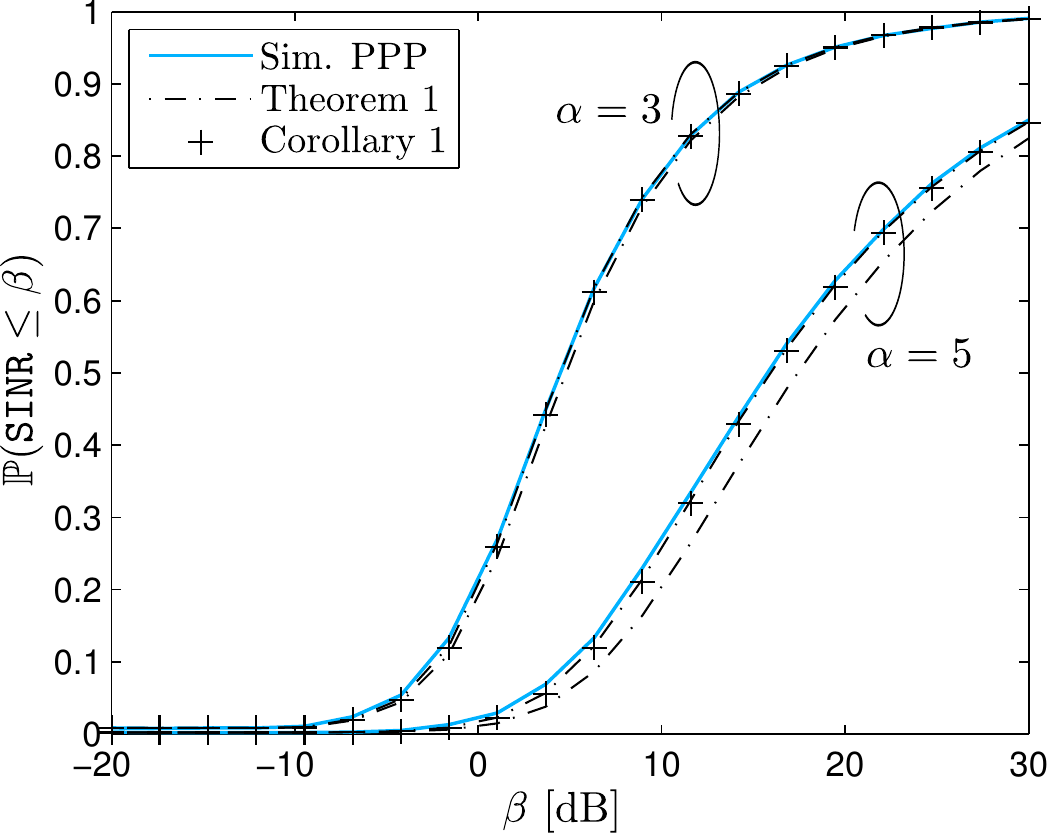}
	\label{fig:inter_sir2}}}
	\caption{CDF of $\mathtt{SINR}$. Simulation (solid), bounds from Theorem~\ref{thm:sinr_distribution} (dash-dotted) and approximation from Corollary~\ref{col:approx_sinr} (``+"-marks). $\mathrm{Max}$-$\mathtt{SINR}$ association from \cite{dhillon12} for $\eta\to\infty$ (dotted-diamonds). Non-cooperative nearest-BS association \cite{andrews11} (dashed). Parameters: $\lambda=14\,\text{BS}/\text{km}^2$, $D=300$ m, $\tilde T=0$ dB.}
\end{figure*}

\begin{remark}
The $m$-th derivative in \eqref{eq:sinr_distribution} can be efficiently computed using Fa\`{a} di Bruno's rule \cite{bruno1857} in combination with Bell polynomials \cite{johnson07}, provided the derivatives of the outer and inner function are known.
\end{remark}

The $m$-th derivative of the inner function of $\mathsf{P}$ can be computed in closed-form as shown for \eqref{eq:lap_ppp} next. An equivalent expression for the conditional case can be obtained by setting $\lambda\pi=D^{-2}$.

\begin{lemma}\label{lem:der_laplace_sig}
The $m$-th derivative ($m>0$) of the exponent of $\lap_{\sig}(-s)$ at $s=-\frac{1}{\theta\beta}$ is given by
\begin{IEEEeqnarray}{rCl}
\tfrac{2}{\alpha}\lambda\pi (\theta\beta)^{m-\frac{2}{\alpha}}\mathbb{E}\left[\mathsf{g}^{\frac{2}{\alpha}}\Gamma(m-\tfrac{2}{\alpha},\tfrac{\mathsf{g}}{\theta\beta}\max\{D^{-\alpha},\hspace{-.02cm}\tfrac{T}{\mathsf{g}}\})\right].\IEEEeqnarraynumspace
\end{IEEEeqnarray}
\end{lemma}
\begin{IEEEproof}
See Appendix~\ref{ap:der_laplace_sig}.
\end{IEEEproof}

\begin{figure*}[!t]
	\centerline{\subfloat[$\mathtt{SINR}$ CDF (unconditional case)]{\includegraphics[width=0.488\textwidth]
	{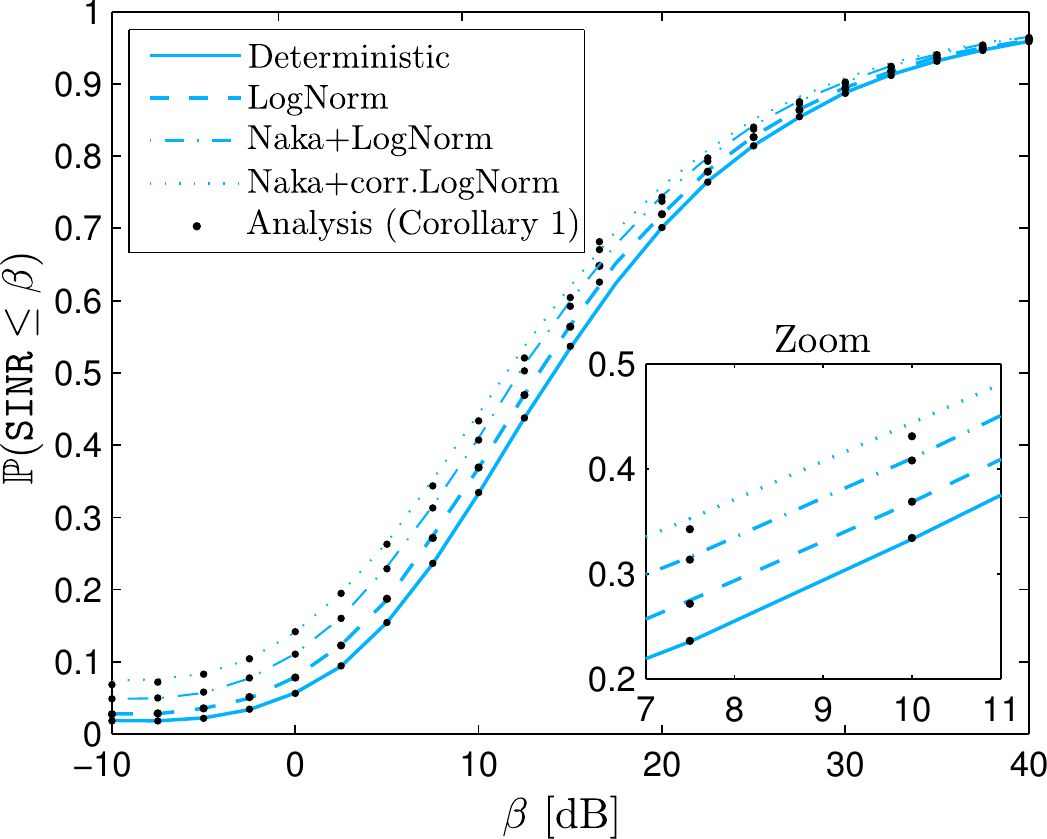}
	\label{fig:inter_sir_fading}}
	\hfil
	\subfloat[Spectral Efficiency $\mathtt{R}$ CDF (unconditional case)]{\includegraphics[width=0.485\textwidth]
	{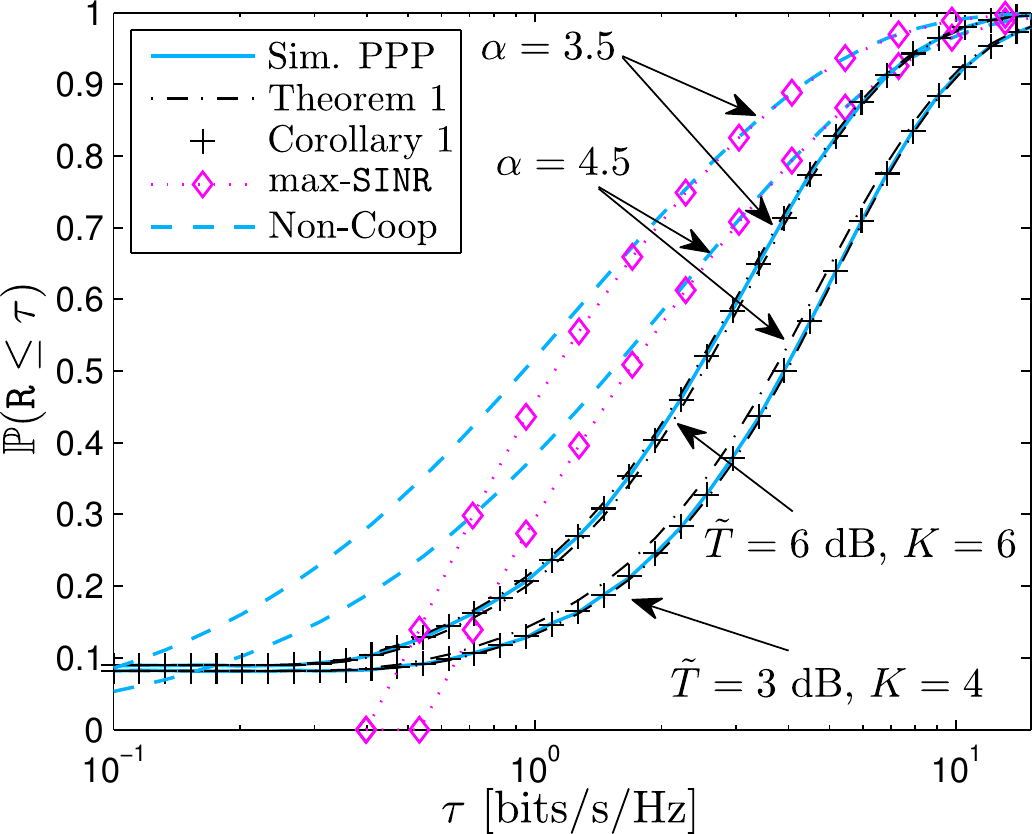}
	\label{fig:inter_se}}}
	\caption{(a) CDF of $\mathtt{SINR}$ for different fading distributions. Deterministic (no fading, $\mathsf{g}\equiv 1$) (solid line), Lognormal shadow fading with standard deviation $\sigma_{\text{dB}}=6$ (dashed), composite Nakagami+Lognormal fading with Nakagami parameter $m=4$ and $\sigma_{\text{dB}}=8$ (dashed-dotted), and composite Nakagami with correlated Lognormal fading with Nakagami parameter $m=2$ and $\sigma_{\text{dB}}=10$ (dotted). Marks represent analytical results (Corollary~\ref{col:approx_sinr}). (b) CDF of $\mathtt{R}$ for exponential fading. Simulation (solid), bounds from Theorem~\ref{thm:sinr_distribution} (dash-dotted) and approximation from Corollary~\ref{col:approx_sinr} (``+"-marks). $\mathrm{Max}$-$\mathtt{SINR}$ association from \cite{dhillon12} for $\eta\to\infty$ (dotted+diamonds). Non-cooperative nearest-BS association \cite{andrews11} (dashed). Parameters: $\lambda=14\,\text{BS}/\text{km}^2$, $\eta=162$ dB, $\alpha=4.5$.}
\end{figure*}

\subsection{Numerical Examples}
Fig.~\ref{fig:inter_sir1} shows the empirical CDF of the $\mathtt{SINR}$ together with the theoretical results (Theorem~\ref{thm:sinr_distribution} and Corollary~\ref{col:approx_sinr}) for the unconditional case and different values of $\alpha$. The transmit SNR was set to $\eta=162$ dB and the fading gains were assumed to follow a unit-mean exponential distribution (Rayleigh fading). The value $D=300$ m corresponds to $3$ cooperating BSs on average. It can be seen that the Gamma approximation from Proposition~\ref{prop:interference_ap} is accurate. Also, the gap between the lower and upper bound enclosing the estimated CDF is fairly small, but it increases for larger $\alpha$. Finally, the simple approximation from Corollary~\ref{col:approx_sinr} performs remarkably well. For comparison, the CDF of the $\mathtt{SINR}$ with instantaneous $\max$-power association from \cite{dhillon12}, which models DCS/TPS cooperation, is also plotted (CDF accurate for $\beta>-4$ dB). It can be seen that aggressively turning interference into useful signal leads to a higher $\mathtt{SINR}$ than with the $\max$-power association. However, as NC-JT consumes more radio resources, the net gain for highly loaded cells may not be in favor of NC-JT. The performance for non-cooperative downlink transmission with \emph{average} $\max$-power cell association from \cite{andrews11} is also shown for reference. 

Similarly, Fig.~\ref{fig:inter_sir2} shows the results for the conditional case with $K=3$ cooperating BSs ($K=\lambda\pi D^2$ with $\lambda,D$ as in Fig.~\ref{fig:inter_sir1}) and the same parameters. In contrast to the unconditional case, we now observe that for larger $\alpha$ the Gamma approximation from Proposition~\ref{prop:interference_ap} slightly looses accuracy. This is due to the aforementioned negative correlation, which comes into effect at larger $\alpha$ since intra-cluster interference $\mathsf{J}_{\mathcal{C}}$ then dominates out-of-cluster interference $\mathsf{J}_{\mathcal{\bar C}}$.

Given the fact that the Gamma approximation cannot in general preserve the true interference tail behavior for all distributions of $\mathsf{g}$ (cf. Remark~\ref{rem:fading}), it is important to compare analysis and simulation for different fading statistics. Fig.~\ref{fig:inter_sir_fading} shows the $\mathtt{SINR}$ CDF for three different assumptions about the fading distribution, namely deterministic (or no fading), Lognormal shadowing and Nakagami-$m$ with Lognormal shadowing (with and without correlation). A correlation model similar to\cite{klingenbrunn99} was used for creating correlated Lognormal random variables. The analytical results are shown for the approximation from Corollary~\ref{col:approx_sinr}. It can be seen that the Gamma approximation leads to a small deviation from the true $\mathtt{SINR}$ CDF. However, for fading distributions with a considerably different tail, e.g., Lognormal shadowing, which has a heavy-tailed distribution, this bias becomes perceptible. In particular for Nakagami fading plus correlated Lognormal shadowing with large standard deviation, the analytical results are slightly biased.

Fig.~\ref{fig:inter_se} shows the CDF of $\mathtt{R}$ for different system parameters. It can be seen that the accuracy of Theorem~\ref{thm:sinr_distribution} and Corollary~\ref{col:approx_sinr} is not affected by the transformation $\mathtt{R}=\log_{2}(1+\mathtt{SINR})$.

\section{Application of the Main Result}\label{sec:application}
In the following, the developed model is used to further investigate the inherent trade-offs of NC-JT.
\subsection{Effect of Imperfect CSI-R on NC-JT Cooperation}\label{sec:csi}
While coarse CSI-T may be already sufficient for deciding whether a user should obtain cooperation from a BS, higher requirements on the accuracy are imposed on the CSI-R; at the receiver, the \emph{composite} channel subsuming all individual cooperative-BS-to-user links must be accurately estimated for coherent detection. Typically, BSs transmit cell-specific \emph{orthogonally-multiplexed} reference signals (RS) to avoid strong inter-RS interference. Due to the inter-RS orthogonality, the channels are estimated independently of each other before they can be combined to yield a final composite estimate. Hence, the final estimate will suffer from estimation-error accumulation. Moreover, the need for orthogonally multiplexing the RSs implies that resources dedicated for channel estimation must be shared among the cooperative BSs, thereby cutting down the per-BS share. It is hence important to characterize the error of the final channel estimate as a function of the number of channels (equivalently, the number of cooperative BSs) to be estimated.

\begin{figure*}[!t]
\centerline{\subfloat[]{
	\includegraphics[width=0.49\textwidth]{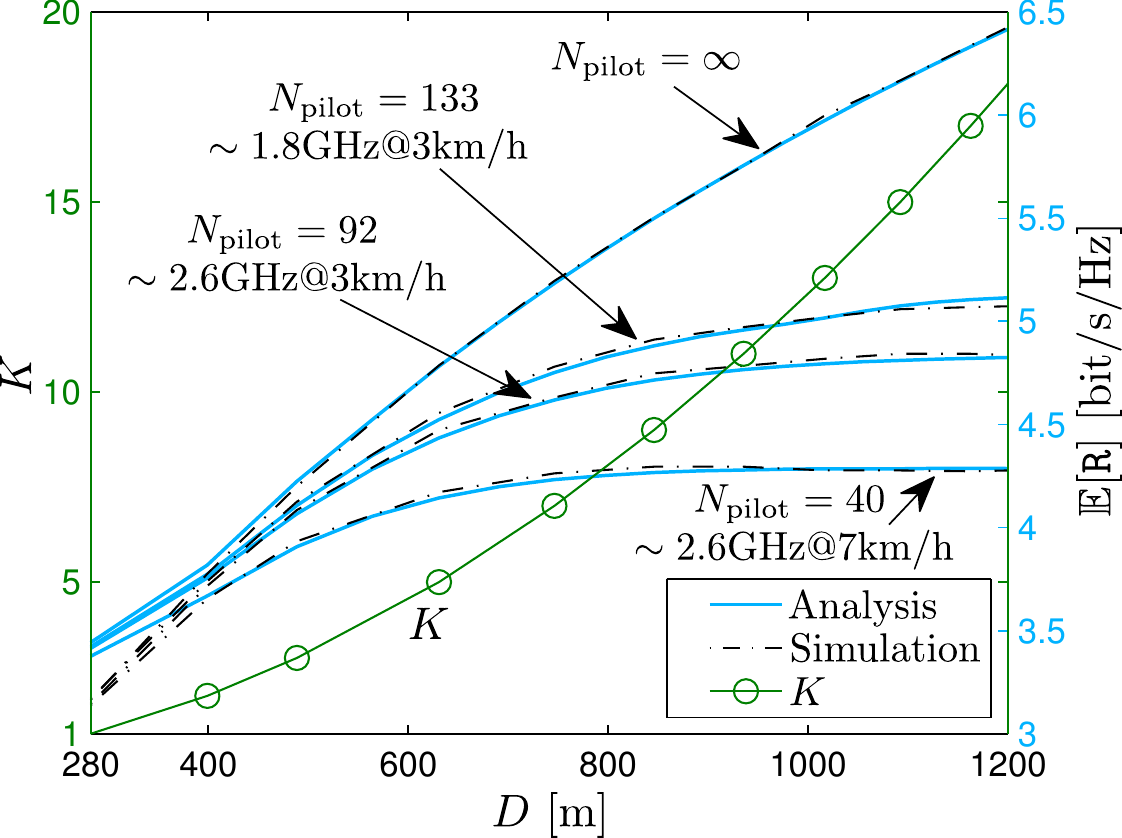}
	\label{fig:inter_opt_cluster_size}}
	\hfil
	\subfloat[]{
	\includegraphics[width=0.466\textwidth]{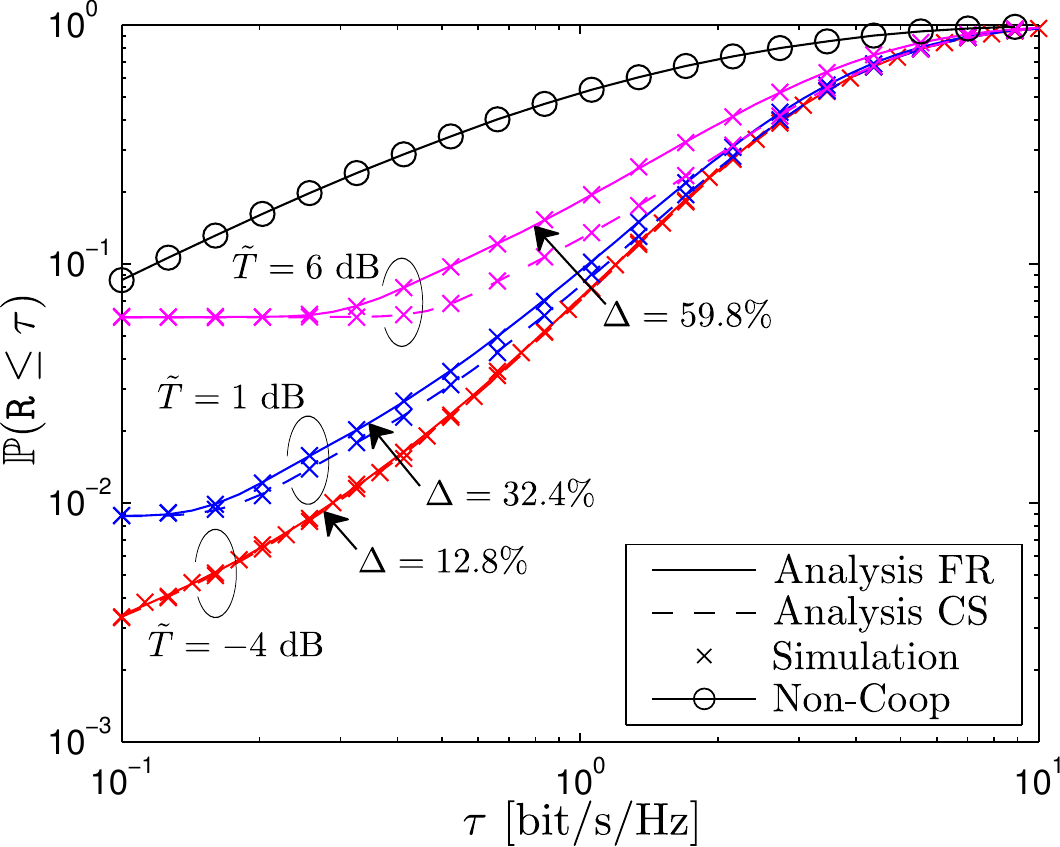}
	\label{fig:se_T_intra}}}
	\caption{(a) Average spectral efficiency $\mathbb{E}[\mathtt{R}]$ vs. the cluster size $K$ for different $N_{\text{pilot}}$. Parameters are: BS density $\lambda=4\,\text{BS}/\text{km}^2$, $\alpha=4$, $T=0$, $\eta=162$ dB. Unit-mean exponential fading. (b) CDF of $\mathtt{R}$ for different $T$ for the unconditional case. Approximation from Corollary~\ref{col:approx_sinr} for FR (solid) and CS (dashed). Simulation with Poisson interference (``x"-marks). Parameters are: $\alpha=3.5$, $\lambda=14\,\text{BS}/\text{km}^2$, $\eta=162$ dB. Average number of cooperative BSs $K=7$ ($D=400$~m). Perfect CSI assumed.}
\end{figure*}

For pilot-based channel estimation under MMSE criterion, the effect of imperfect CSI-R can be captured by an equivalent \emph{effective} $\mathtt{SINR}$  \cite{hassibi03}; strictly speaking, the receive-filter mismatch reduces the useful signal power while causing residual interference. The MMSE of the $i$-th channel estimate (to the $i$-th cooperative BS) given $K$ transmitters (BSs) has the form \cite{hassibi03}
\begin{IEEEeqnarray}{rCl}
	\sigma_{\text{MMSE},i}^2 &=& \frac{1}{1+\mathbb{E}_{\mathsf{g}_{i}}\left[\mathtt{SINR}_{\text{pilot},i}\right]\frac{N_{\text{pilot}}}{K}},\IEEEeqnarraynumspace
\end{IEEEeqnarray}
where $\mathbb{E}_{\mathsf{g}_{i}}\left[\mathtt{SINR}_{\text{pilot}}\right]$ and $N_{\text{pilot}}$ are the \emph{fading-averaged} pilot-signal-to-interference-plus-noise ratio and the total number of sampling points (pilot symbols) for channel estimation during one channel coherence period, respectively, associated with the $i$-th cooperative BS. The factor $1/K$ accounts for the aforementioned need for sharing the resources dedicated to channel estimation among the cooperative BSs. Assuming the same transmit power for all pilot symbols and noting that inter-RS interference is avoided inside the cluster, $\sigma_{\text{MMSE},i}^2$ can be further expressed as
\begin{IEEEeqnarray}{rCl}
	\sigma_{\text{MMSE},i}^2&=&\frac{1}{1+\frac{\|\mathsf{x}_{i}\|^{-\alpha}}{\mathsf{J}_{\mathcal{\bar C}}+\frac{1}{\eta}}\frac{N_{\text{pilots}}}{K}},
		\label{eq:mmse_i}
\end{IEEEeqnarray}
where $\mathsf{J}_{\mathcal{\bar C}}+\tfrac{1}{\eta}$ is the effective estimation noise. Noting that the estimation error accumulates with $K$, the resulting effective $\mathtt{SINR}$ can then we written similarly to \cite{lozano12} as
\begin{IEEEeqnarray}{rCl}
	\mathtt{SINR}=\frac{\sum\limits_{i\in\Phi\cap\mathcal{C}}\hspace{-.03cm}(1-\sigma_{\text{MMSE},i}^2)\,\mathsf{g}_{i}\|\mathsf{x}_{i}\|^{-\alpha}\mathds{1}(\mathsf{g}_{i}\|\mathsf{x}_{i}\|^{-\alpha}\geq T)}{
		\mathsf{J}_{\text{CSI}}+\mathsf{J}_{\mathcal{C}}+\mathsf{J}_{\mathcal{\bar C}}+\frac{1}{\eta}},\label{eq:sinr_csi}\IEEEeqnarraynumspace
\end{IEEEeqnarray}
where $\mathsf{J}_{\text{CSI}}\mathdef\sum_{i\in\Phi\cap\mathcal{C}}\sigma_{\text{MMSE},i}^2\,\mathsf{g}_{i}\|\mathsf{x}_{i}\|^{-\alpha}\mathds{1}(\mathsf{g}_{i}\|\mathsf{x}_{i}\|^{-\alpha}\geq T)$ is the residual interference due to imperfect CSI-R. Unfortunately, two problems arise from \eqref{eq:sinr_csi}, which would make an exact analysis cumbersome: first, the useful received power and the residual interference are statistically dependent through $\sigma_{\text{MMSE},i}^2$; second, the latter two quantities in turn, depend on $\mathsf{J}_{\mathcal{\bar C}}$ since $\sigma_{\text{MMSE},i}^2$ is a function of $\mathsf{J}_{\mathcal{\bar C}}$. In order to circumvent this intractability, we propose the following two-step approximation: 1) We replace the out-of-cluster interference $\mathsf{J}_{\mathcal{\bar C}}$ in \eqref{eq:mmse_i} by $\mathbb{E}\left[\mathsf{J}_{\mathcal{\bar C}}\right]$, which can be computed using Campbell's theorem\cite{HaenggiBook}, cf. Appendix~\ref{ap:interference_ap}; 2) Similar to the conditional case in Section~\ref{sec:bpp}, we further model the useful received power and the residual interference as being statistically independent. The residual interference is then incorporated in the Gamma random variable $\mathsf{\tilde J}$, for which the shape $k$ and scale $\theta$ must be re-computed using the second-order moment-matching technique already used in Proposition~\ref{prop:interference_ap}. This requires calculating the mean and variance of $\mathsf{J}_{\text{CSI}}$, which for the conditional case yields
\begin{IEEEeqnarray}{rCl}
\mathbb{E}\left[\mathsf{J}_{\text{CSI}}\right]&=&\mathbb{E}\left[\sum\limits_{i=0}^{K-1}\sigma_{\text{MMSE},i}^2\,\mathsf{g}_{i}\|\mathsf{x}_{i}\|^{-\alpha}\,\mathds{1}(\mathsf{g}_{i}\|\mathsf{x}_{i}\|^{-\alpha}\geq T)\right]\IEEEnonumber\\
&=&K\hspace{-.02cm}\times\hspace{-.02cm}\frac{1}{b_KD^{2}}\,\mathbb{E}\Big[\mathsf{g}\min\left\{D^\alpha,\tfrac{\mathsf{g}}{T}\right\}^{\frac{2}{\alpha}}\IEEEnonumber\\
&&\qquad\hspace{-.03cm}\underbrace{\times{}_2F_{1}\hspace{-.07cm}\left(1,\tfrac{2}{\alpha},1+\tfrac{2}{\alpha};-b_K^{-1}\min\left\{D^\alpha,\tfrac{\mathsf{g}}{T}\right\}\right)\Big]}_{\mathdef \mu_K}\IEEEeqnarraynumspace
\end{IEEEeqnarray}
and
\begin{IEEEeqnarray}{rCl}
\text{Var}\left[\mathsf{J}_{\text{CSI}}\right]&=&\mathbb{E}\left[\sum\limits_{i=0}^{K-1}\sigma_{\text{MMSE},i}^4\mathsf{g}_{i}^2\|\mathsf{x}_{i}\|^{-2\alpha}\,\mathds{1}(\mathsf{g}_{i}\|\mathsf{x}_{i}\|^{-\alpha}\geq T)\right]\IEEEnonumber\\
&&\quad-K\mu_K^2\IEEEnonumber\\
&=&-K\mu_K^2+\frac{K}{b_K^2D^2}\,\mathbb{E}\Big[\mathsf{g}^2\min\left\{D^\alpha,\tfrac{\mathsf{g}}{T}\right\}^{\frac{2}{\alpha}}\IEEEnonumber\\
&&\quad\times{}_2F_{1}\hspace{-.07cm}\left(2,\tfrac{2}{\alpha},1+\tfrac{2}{\alpha};-b_K^{-1}\min\left\{D^\alpha\hspace{-.07cm},\tfrac{\mathsf{g}}{T}\right\}\right)\Big],\IEEEeqnarraynumspace
\end{IEEEeqnarray}
 where $b_K\mathdef N_{\text{pilots}}/K/(\mathbb{E}\left[\mathsf{J}_{\mathcal{\bar C}}\right]+\tfrac{1}{\eta})$ and ${}_2F_{1}(a,b,c;z)$ is the Gaussian hypergeometric function\cite{olver10}. Averaging over $K$ gives the corresponding values for the unconditional case. The Laplace transform of the numerator in \eqref{eq:sinr_csi} for the conditional case is given by
\begin{IEEEeqnarray}{rCl}
	\left(\frac{2}{\alpha\,D^2}\int_{D^{-\alpha}}^{\infty}t^{-1-\frac{2}{\alpha}}\,\mathcal{L}_{\mathsf{g}\mathds{1}(\mathsf{g}\geq T/t)}\left(\frac{s\,t^2}{t+1/b_K}\right)\,\mathrm dt\right)^{K}.\IEEEeqnarraynumspace
\end{IEEEeqnarray}
For $\mathsf{g}\sim\text{Exp}(1)$, we have
\begin{IEEEeqnarray}{rCl}
\mathcal{L}_{\mathsf{g}\mathds{1}(\mathsf{g}\geq T/t)}\left(u\right)&=&1-e^{-\frac{T}{t}}+\frac{e^{-\frac{T}{t}\left(u+1\right)}}{u+1}.\IEEEeqnarraynumspace
\end{IEEEeqnarray}
See Appendix~\ref{ap:laplace_sig_ppp} for obtaining the corresponding expression for the unconditional case. 

The approximation 2) is justified by the following observation: looking at the denominator of \eqref{eq:sinr_csi}, we see that BSs that would notably contribute to the residual interference because of a high MMSE must be located farther away from the typical user. Because of their potentially large distance to the typical user, these BSs experience a high path loss, and hence by the cooperation activation policy are unlikely to serve the typical user. As will be shown later, the deviation from the true $\mathtt{SINR}$ caused by these approximations remains fairly small.
\begin{remark}
	An alternative way for studying the impact of imperfect CSI on the system performance is to keep the level of acceptable CSI error constant while increasing the spectrum overhead, i.e., the number of pilot symbols $N_{\text{\textnormal{pilots}}}$. This eventually leads to pilot contamination, which similarly limits the obtainable throughput. 
\end{remark}

\textit{Optimal cluster size under imperfect CSI-R:}
We will focus on the conditional case ($\Phi(\mathcal{C})=K$) in the following and analyze the impact of imperfect CSI-R for different cluster sizes $K$. In order to apply Theorem~\ref{thm:sinr_distribution} to the $\mathtt{SINR}$ in \eqref{eq:sinr_csi}, the Laplace transform of the numerator must be calculated, in addition to obtaining the moments of $\mathsf{J}_{\text{CSI}}$ for re-computing $k$ and $\theta$. We skip these tasks since they are technically the same as for Proposition~\ref{prop:interference_ap}, Lemma~\ref{lem:laplace_sig_bpp} and Lemma~\ref{lem:der_laplace_sig}. Fig.~\ref{fig:inter_opt_cluster_size} shows the average spectral efficiency $\mathbb{E}\left[\mathtt{R}\right]$ vs. $K$ for different $N_{\text{pilot}}$. The BS density was set to $\lambda=4\,\text{BS}/\text{km}^2$. The values for $N_{\text{pilot}}$ were chosen according to the LTE channel estimation specifications \cite{gosh10}. The average spectral efficiency was obtained using the relation $\mathbb{E}[\mathtt{R}]=\int_{0}^{\infty}\mathbb{P}(\mathtt{SINR}>2^{\tau}-1)\,\mathrm d\tau$ and using Corollary~\ref{col:approx_sinr}. It can be seen that, as expected, the average spectral efficiency runs into saturation for large $K$. In contrast, imperfect CSI-R has only little effect on the spectral efficiency in the small $K$ regime, where out-of-cluster interference $\mathsf{J}_{\mathcal{\bar C}}$ limits the performance; here, increasing $K$ does not change the MMSE much. As can be seen, the saturation point is roughly around $K=7$ for typical $N_{\text{pilot}}$.

\begin{remark}
Note that the saturation point in terms of throughput may be even smaller due to the cell-load increase induced by NC-JT. When cell load is the limiting factor, e.g., in high-load scenarios, the per-user throughput may even strictly decrease with the cluster size. It is hence likely that considerably larger cluster sizes will not be beneficial, irrespective of the engineering effort spent on the backhaul.
\end{remark}
Furthermore, it can be seen that the approximation explained above is noticeable only at $K=1$, whereas it provides a good fit to the simulation results over the whole range of $K$. The effect of $T$ is only marginal and is therefore not shown. Note that the actual saturation trend described in Fig.~\ref{fig:inter_opt_cluster_size} applies to the NC-JT scheme considered in this work, however, similar trends were observed for instance in\cite{lozano12,mondal12} for other cooperation schemes.

\subsection{NC-JT Intra-Cluster Scheduling: FR vs. CS}\label{sec:scheduling}
As mentioned in Section~\ref{sec:model}, in NC-JT cooperating BSs that do not participate in an ongoing joint transmission can act in two different ways: 1) they can reuse the radio resources on which NC-JT is performed (intra-cluster FR). This scheme has been assumed throughout this work; 2) since reusing the same radio resources may translate to an unacceptable degree of intra-cluster interference, the ones used for NC-JT may alternatively be prohibited for other transmissions (intra-cluster CS). Even though intra-cluster interference is now effectively avoided, the overall benefit of CS is not obvious; exclusively reserving radio resources for joint transmission virtually increases the cell load at the cooperating BSs that do not participate in an ongoing joint transmission. Such a load increase may be unacceptable as other users may possibly suffer from radio resources shortage, and hence may experience lower data rates. Quantifying this trade-off and comparing these two scheduling schemes requires an accurate description of the different system components, among which the cooperation activation mechanism has a major impact.   

Instead of trying to precisely characterize the effect of NC-JT on the actual cell load ---which is difficult in a cooperative multi-cell scenario, cf. \cite{singh13} for a possible approach--- an alternative way that captures the first-order relative behavior of the two scheduling schemes is chosen: from the perspective of NC-JT with CS, switching to FR invokes a resource saving at the cooperative BSs not involved in a joint transmission since the resources used for joint transmission can be reused. This saving directly translates into a load reduction at these BSs.

Although in practice the choice between FR and CS may be made according to the actual BS locations and load situation, we next study the average radio resource saving to reveal the underlying trend. Hence, we define
\begin{IEEEeqnarray}{rCl}
	\Delta\mathdef1-\mathbb{E}\left[\frac{\sum\limits_{i=\Phi\cap\mathcal{C}}\mathds{1}\left(\mathsf{g}_{i}\|\mathsf{x}_{i}\|^{-\alpha}\geq T\right)}{\sum\limits_{i\in\Phi}\mathds{1}(\mathsf{x}_{i}\in\mathcal{C})}\right],\IEEEeqnarraynumspace\label{eq:delta}
\end{IEEEeqnarray}
which describes the \emph{spatially}-averaged radio resource saving when switching from CS to FR. Applying the law of the iterated expectation, $\Delta$ can be computed as 
\begin{IEEEeqnarray}{rCl}
	\Delta&=&1-\mathbb{E}_{K}\left[\frac{1}{K}\mathbb{E}\left[\sum\limits_{i=0}^{K-1}\mathds{1}(\mathsf{g}_{i}\|\mathsf{x}_{i}\|^{-\alpha}\geq T)\right]\right]\IEEEnonumber\\
	&=&1-\mathbb{E}\left[\min\left\{1,\tilde T^{-\frac{2}{\alpha}}\mathsf{g}^{\frac{2}{\alpha}}\right\}\right].\IEEEeqnarraynumspace\label{eq:load_reduction_general}
\end{IEEEeqnarray}
Interestingly, $\Delta$ is independent of $\lambda$. For $\mathsf{g}~\sim\text{Exp}(1)$, \eqref{eq:load_reduction_general} reduces to $\Delta=1-\exp(-\tilde T)-\tilde T^{-2/\alpha}\gamma(1+\tfrac{2}{\alpha},\tilde T).$

Fig.~\ref{fig:se_T_intra} shows the CDF of $\mathtt{R}$ for NC-JT with the two scheduling policies FR and CS. First, it can be seen that the $\mathtt{SINR}$ improves as $T$ decreases (cooperation is aggressively triggered). Consequently, the number of consumed radio resources are at highest for small $T$. Here, switching from CS to FR does barely change the $\mathtt{SINR}$ statistics while a radio resource saving of approximately $12.8$~\% is achieved. In this regime, FR may thus be more favorable. For larger $T$ one has to bite the bullet: much higher savings, e.g., up to $60$\% in Fig.~\ref{fig:se_T_intra}, can be achieved, however, at the cost of worsening the $\mathtt{SINR}$ due to higher intra-cluster interference. In moderately loaded cells, FR may be mandatory for NC-JT to not overload cells. In contrast, CS should be used in lightly loaded cells to additionally profit from muting intra-cluster interference.

\section{Discussion and Conclusion}\label{sec:conclusion}
We developed a tractable model for analyzing NC-JT BS cooperation. We characterized the $\mathtt{SINR}$ CDF for a typical user for the case of user-centric BS clustering. This result is tractable and fairly general, for instance no specific fading distribution is assumed. It was found that the gains of cooperation increase with the path loss exponent. Also, uniformly increasing the BS density while fixing the cooperation radius improves the $\mathtt{SINR}$. Furthermore, the average spectral efficiency was shown to saturate at a cluster size of around $7$ BSs when CSI-R is imperfect. Complementing earlier work, this result provides insights for practical system design. We showed that for NC-JT, intra-cluster CS should be used in lightly loaded cells with generous channel-dependent cooperation activation, while intra-cluster FR should be used otherwise.

Although the model developed in this work led to new insights regarding the performance of NC-JT in lightly loaded cells, it possesses some shortcomings that could be addressed in future work. For instance, including user-centric BS clustering that is based on the RSS \emph{difference} to the serving BS (in contrast to assuming a fixed cooperation radius) would yield a more practical model, where only cell-edge user receive cooperation. Capturing cell load and resource allocation inter-cell dependencies would further contribute to a better understanding of the trade-offs involved in NC-JT, especially when looking beyond lightly loaded cells. Since this work focuses on the cell-average performance, modifying the model to quantify the cell-edge performance would be interesting as well. Finally, a comprehensive spatial model for analyzing coherent JT is also still not available in the literature.




%

\appendix

\section{~}\label{ap:proofs}
\subsection{Non-Coherent JT: Transmission, Reception and \texorpdfstring{$\mathtt{SINR}$}{SINR}}\label{sec:nc_jt_scheme}
This section explains the transmission/reception procedure under NC-JT and derives the resulting $\mathtt{SINR}$ used in \eqref{eq:sinr_nc}. In an OFDMA-based cellular network, BSs transmit a complex-valued OFDM signal that is prepended by a cyclic prefix (CP). By letting the channel appear to provide a circular convolution, the CP ensures that timely-dispersed multipath versions of the useful signal do not create inter-symbol interference\cite{gosh10}. NC-JT exactly makes use of this property. Consider $K$ cooperative BSs jointly serving a user, for which the received discrete time-domain signal (OFDM symbol) can be written as
\begin{IEEEeqnarray}{rCl}
	r[n]&=&\sum\limits_{k=0}^{K-1}(h_{k}\circledast s)[n]+i[n]+z[n],\quad n=1,\ldots,N,\IEEEeqnarraynumspace
\end{IEEEeqnarray}
where $s[n]$, with $\mathbb{E}[|s[n]|^2]=\rho$, is the useful signal transmitted by all $K$ cooperating BSs, $i[n]$ is the sum interference signal from all other BSs, $z[n]$ is the receiver noise modeled as circular-symmetric complex AWGN with variance $\sigma^2$ and $N$ is the OFDM symbol length (discrete Fourier transform (DFT) length). The symbol ``$\circledast$" refers to the circular convolution, i.e., $(h_{k}\circledast s)[n]\mathdef\sum_{\ell=0}^{N-1}h_{k}[\ell]\,s[(n-\ell)\mod N]$ induced by the CP. The {\it effective} time-domain channel to the $k$-th cooperating BS is given by $h_{k}[n]$ and accounts for average path loss, channel fading, and, in addition, the timing offset due lack of tight BS synchronization in NC-JT. Hence, we write $h_{k}[n]$ as
\begin{IEEEeqnarray}{rCl}
	h_{k}[n]=a_{k}[n-\nu_k]\,\|x_{k}\|^{-\alpha/2},\label{eq:nc_jt_scheme1}
\end{IEEEeqnarray}
where $a_{k}[n]$ is the complex-valued time-domain fading coefficient characterizing the channel fading and $\nu_{k}$ is the additional timing offset at the $k$-th cooperating BS. Due to lack of tight coordination, it is reasonable to treat the $\nu_k$s as being independent across BSs. Similarly, the $a_{k}$s can be assumed independent across BSs since no phase-mismatch correction is performed by the cooperative BSs in NC-JT. After serial-parallel conversion, $r[n]$ is transformed into frequency domain by the DFT. Provided the CP length is properly chosen, the equivalent frequency-domain signal is
\begin{IEEEeqnarray}{rCl}
R[m]	&=&	\text{DFT}\left\{\sum\limits_{k=0}^{K-1}(h_{k}\circledast s)[n]+i[n]+z[n]\right\}\IEEEnonumber\\
		&=&	\sum\limits_{k=0}^{K-1}H_{k}[m]\,S[m]+I[m]+Z[m].\IEEEeqnarraynumspace
\end{IEEEeqnarray}
With \eqref{eq:nc_jt_scheme1}, $H_{k}[m]$ is computed as
\begin{IEEEeqnarray}{rCl}
H_{k}[m]	&=&	\frac{1}{\sqrt{N}}\sum\limits_{n=0}^{N-1}h_{k}[n]\,e^{-j2\pi m\frac{n}{N}}\IEEEnonumber\\
			&=&\sqrt{g_{k}[m]}\,\|x_{k}\|^{-\alpha/2}\,e^{j\phi_{k}[m]+j2\pi m\frac{\nu_{k}}{N}},
\end{IEEEeqnarray}
where $\sqrt{g_{k}}$ is the (frequency-domain) fading envelope of the power fading gain $g_{k}$ introduced in Section~\ref{sec:model} and $\phi_{k}$ is the corresponding phase rotation. The additional modulation term $e^{j2\pi m\nu_{k}/N}$ is a result of the DFT time-shift property. The received useful signal on subcarrier $m$ then has the form
\begin{IEEEeqnarray}{c}
	R_{\text{use}}[m]=S[m]\sum\limits_{k=0}^{K-1}\sqrt{g_{k}[m]}\,\|x_{k}\|^{-\alpha/2}e^{j\phi_{k}[m]+j2\pi m\frac{\nu_{k}}{N}},\label{eq:nc_jt_scheme2}\IEEEeqnarraynumspace
\end{IEEEeqnarray}
The sum in \eqref{eq:nc_jt_scheme2} can be seen as the effective channel gain on subcarrier $m$. With CSI-R, the corresponding complex phase of this gain is corrected and the useful signal power becomes
\begin{IEEEeqnarray}{rCl}
&&\left\lvert R_{\text{use}}[m]\right\lvert^2=|S[m]|^2\sum\limits_{k=0}^{K-1}g_{k}[m]\|x_{k}\|^{-\alpha}\IEEEnonumber\\
&&\quad+|S[m]|^2\sum\limits_{\substack{k,\ell=0\\k\neq\ell}}^{K-1}\hspace{-.1cm}\sqrt{\tfrac{g_{k}[m]g_{\ell}[m]}{\|x_{k}\|^{\alpha}\|x_{\ell}\|^{\alpha}}}\,\mathcal{R}\left[e^{j\tilde\phi_{k\ell}[m]}\,e^{j2\pi m\frac{\tilde\nu_{k\ell}}{N}}\right],\IEEEeqnarraynumspace
\end{IEEEeqnarray}
where $\tilde\phi_{k\ell}[m]\mathdef \phi_k[m]-\phi_\ell[m]$, $\tilde\nu_{k\ell}\mathdef\nu_k-\nu_\ell$ and $\mathcal{R}(\cdot)$ denotes the real part. Note that while the $g_k$s and $\phi_k$s remain constant within the coherence bandwidth of the fading channel (usually, a few tens times the subcarrier spacing), this may not be the case for the modulation term $\exp(j2\pi m\nu_{k}/N)$, which varies in $m$ with frequency $\nu_k/N$. For instance, when $\nu_k$ corresponds to half the (extended) cyclic prefix duration, $\exp(j2\pi m\nu_{k}/N)$ has a period of only $8$ subcarriers, assuming a 10 MHz LTE system\cite{gosh10}. When both the $\tilde\nu_k$s and the coherence bandwidth are relatively large, the modulation term causes the received signal power to vary considerably over a large number of subcarriers having the same $g_k[m]=g_k$ and $\phi_k[m]=\phi_k$. To capture the overall effect of the timing offset, it is hence reasonable to average over these power variations within the coherence bandwidth (spanning $N_{\text{c}}$ subcarriers around subcarrier $m$), i.e.,
\begin{IEEEeqnarray}{rCl}
	\overline{|R_{\text{use}}|^2}&=&\sum\limits_{k=0}^{K-1}g_{k}\|x_{k}\|^{-\alpha}\,\frac{1}{N_{\text{c}}}\sum\limits_{u=0}^{N_{\text{c}}-1}|S[u]|^2\IEEEnonumber\\
&&\hspace{-1.2cm}+\sum\limits_{\substack{k,\ell=0\\k\neq\ell}}^{K-1}\hspace{-.1cm}\frac{\sqrt{g_{k}g_{\ell}}}{\|x_{k}\|^{\alpha}\|x_{\ell}\|^{\alpha}}\,\frac{1}{N_{\text{c}}}\sum\limits_{u=0}^{N_{\text{c}}-1}|S[u]|^2\,\mathcal{R}\left[e^{j\tilde\phi_{k\ell}}\,e^{j2\pi u \frac{\tilde\nu_{k\ell}}{N}}\right],\IEEEnonumber\\
&\approx&\frac{\rho}{N}\sum\limits_{k=0}^{K-1}g_{k}\|x_{k}\|^{-\alpha},\label{eq:nc_jt_scheme3}
\end{IEEEeqnarray}
where we used the fact that $\tfrac{1}{N_{\text{c}}}\sum_{u=0}^{N_{\text{c}}-1}|S[u]|^2\approx\mathbb{E}[s[n]]/N=\rho/N$ (assuming an even power allocation across the $N$ subcarriers) and that $\tfrac{1}{N_{\text{c}}}\sum_{u=0}^{N_{\text{c}}-1}|S[u]|^2\mathcal{R}[e^{j\tilde\phi_{k\ell}}e^{2\pi u \tilde\nu_{k\ell}/N}]\approx0$ for large $N_{\text{c}}$. Similarly, the interference-plus-noise power (signal variance) can readily be obtained as
\begin{IEEEeqnarray}{rCl}
	\overline{\left\lvert I+Z\right\lvert^2}=\frac{\rho}{N}\sum\limits_{\substack{\text{interfering}\\ x_k}}g_{k}\|x_{k}\|^{-\alpha}+\frac{\sigma^2}{N},\IEEEeqnarraynumspace\label{eq:nc_jt_scheme4}
\end{IEEEeqnarray}
where we have exploited the fact that interfering signals transmitted by other (possibly cooperating) BSs superimpose non-coherently at the receiver. Combining \eqref{eq:nc_jt_scheme3} and \eqref{eq:nc_jt_scheme4} and defining $\eta\mathdef\rho/\sigma^2$ yields the $\mathtt{SINR}$ expression in \eqref{eq:sinr_nc}.

\subsection{Proof of Proposition~\ref{prop:interference_ap}}\label{ap:interference_ap}
The parameters $k$ and $\theta$ satisfy the relations $\mathbb{E}[\mathsf{\tilde J}]=k\theta$ and $\text{Var}[\mathsf{\tilde J}]=k\theta^2$ \cite{feller71}. The moments of $\mathsf{\tilde J}$ can be computed using Campbell's theorem \cite{stoyan95} as
\begin{IEEEeqnarray}{rCl}
\mathbb{E}[\mathsf{\tilde{J}}]&=&\frac{1}{\eta}+\mathbb{E}\left[2\pi\lambda\int_{0}^{D}\mathsf{g}\,r^{-\alpha+1}\mathds{1}(\mathsf{g}\,r^{-\alpha}<T)\,\mathrm dr\right]\IEEEnonumber\\
&&+2\pi\lambda\int_{D}^{\infty}\mathbb{E}\left[\mathsf{g}\right]r^{-\alpha+1}\,\mathrm dr\IEEEnonumber\\
	&=&\frac{1}{\eta}+\frac{2\pi\lambda}{\alpha-2}\mathbb{E}\left[\mathsf{g}^2\,\min\{D^{\alpha},\tfrac{T}{\mathsf{g}}\}^{\frac{2}{\alpha}-1}\right]\label{eq:mom_I}
\end{IEEEeqnarray}
and
\begin{IEEEeqnarray}{rCl}
\text{Var}[\mathsf{\tilde{J}}]&=&\mathbb{E}\left[2\pi\lambda\int_{0}^{D}\mathsf{g}^2\,r^{-2\alpha+1}\mathds{1}(\mathsf{g}\,r^{-\alpha}<T)\,\mathrm dr\right]\IEEEnonumber\\
&&+2\pi\lambda\int_{D}^{\infty}\mathbb{E}\left[\mathsf{g}^2\right]r^{-2\alpha+1}\,\mathrm dr\IEEEnonumber\\
		&=&\frac{\pi\lambda}{\alpha-1}\mathbb{E}\left[\mathsf{g}^2\,\min\{D^{\alpha},\tfrac{T}{\mathsf{g}}\}^{\frac{2}{\alpha}-2}\right].\label{eq:var_I}
\end{IEEEeqnarray}
Inserting \eqref{eq:mom_I} and \eqref{eq:var_I} in the above relations and solving for $k$ and $\theta$ yields the result.\qed

\subsection{Proof of Theorem~\ref{thm:sinr_distribution}}\label{ap:sinr_distribution}
By the law of total probability, we can condition $\mathbb{P}(\mathtt{SINR}\leq\beta)$ on $\sig$ yielding
\begin{IEEEeqnarray}{rCl}
\mathbb{P}\left(\sig/\inter\leq\beta\right)&=&\mathbb{E}\left[\mathbb{P}\left(\inter\geq\sig/\beta\right)\right]\IEEEnonumber\\
		&=&\int_{0}^{\infty}\frac{\Gamma\left(k,P/\beta\theta\right)}{\Gamma(k)}\,f_{\sig}(P)\,\mathrm{d}P\IEEEnonumber\\
		&=&\int_{0}^{\infty}\lim_{s\to0}e^{-sP}\,\frac{\Gamma\left(k,P/\beta\theta\right)}{\Gamma(k)}\,f_{\sig}(P)\,\mathrm{d}P\IEEEnonumber\\
		&=&\lim_{s\to0}\int_{0}^{\infty}e^{-sP}\,\frac{\Gamma\left(k,P/\beta\theta\right)}{\Gamma(k)}\,f_{\sig}(P)\,\mathrm{d}P,\IEEEeqnarraynumspace
\end{IEEEeqnarray}
where the last line follows from the dominated convergence theorem. In general, $f_{\sig}$ may have a jump at $P=0$ which corresponds to an initial value $\mathbb{P}(\sig=0)>0$. This would render $f_{\sig}$ non-piecewise-continuous. In view of such a possible jump, we decompose the integral as
\begin{IEEEeqnarray}{rCl}
		&&\lim_{s\to0}\int_{0}^{\infty}e^{-sP}\frac{\Gamma\left(k,P/\beta\theta\right)}{\Gamma(k)}f_{\sig}(P)\,\mathrm{d}P\IEEEnonumber\\
		&&\quad\quad=
		\lim_{P\to0_{-}}\underbrace{\frac{\Gamma\left(k,P/\beta\theta\right)}{\Gamma(k)}}_{=1}f_{\sig}(P)\IEEEnonumber\\
		&&\quad\quad\quad\quad+\lim_{s\to0}\int_{0_{+}}^{\infty}e^{-sP}\frac{\Gamma\left(k,P/\beta\theta\right)}{\Gamma(k)}f_{\sig}(P)\,\mathrm{d}P\IEEEnonumber\\
		&&\quad\quad= f_{\sig}(0_{-})+\lim_{s\to0}\int_{0_{+}}^{\infty}e^{-sP}\,\frac{\Gamma\left(k,P/\beta\theta\right)}{\Gamma(k)}\,f_{\sig}(P)\,\mathrm{d}P.\IEEEeqnarraynumspace
\end{IEEEeqnarray}
Having excluded a possible jump at $P=0$, the PDF of $\sig$ is now strictly continuous. Next, we note that the Laplace transform of $\Gamma\left(k,P/\beta\theta\right)/\Gamma(k)$ is $(1-(1+\theta\beta z)^{-k})/z$ with abscissa of convergence $\sigma_{\Gamma}=-1/\theta\beta$. Similarly, since $f_{\sig}$ is a PDF with non-negative support, it has a Laplace transform with abscissa of convergence $\sigma_{f_{\sig}}=0$ \cite{feller71}. However, since we have excluded a possible jump of $f_{\sig}$ at $P=0$, the corresponding Laplace transform is given by $\lap_\sig(s)-f_{\sig}(0_{-})$. Combining these observations with the fact that $\text{Re}(s)=0>\sigma_{\Gamma}+\sigma_{f_{\sig}}=-1/\theta\beta$, we can apply the $s$-convolution theorem for Laplace transforms \cite{lepage80,zayed96} and write

\begin{IEEEeqnarray}{rCl}
&&f_{\sig}(0_{-})+\lim_{s\to0}\int_{0_{+}}^{\infty}e^{-sP}\,\frac{\Gamma\left(k,P/\beta\theta\right)}{\Gamma(k)}\,f_{\sig}(P)\,\mathrm{d}P\IEEEnonumber\\
&&\quad=f_{\sig}(0_{-})+\lim_{s\to0}\frac{1}{2\pi j}\int_{c-j\infty}^{c+j\infty}\frac{1-(1+\theta\beta z)^{-k}}{z}\IEEEnonumber\\
&&\hspace{4.5cm}\times\left[\mathcal{L}_{\sig}(s-z)-f_{\sig}(0_{-})\right]\,\mathrm dz\IEEEnonumber\\
&&\quad=f_{\sig}(0_{-})+\frac{1}{2\pi j}\int_{c-j\infty}^{c+j\infty}\frac{1-(1+\theta\beta z)^{-k}}{z}\IEEEnonumber\\
&&\hspace{4.5cm}\times\left[\mathcal{L}_{\sig}(-z)-f_\sig(0_{-})\right]\,\mathrm dz\IEEEnonumber\\
&&\quad=f_{\sig}(0_{-})+\underbrace{\frac{1}{2\pi j}\int_{c-j\infty}^{c+j\infty}\frac{1-(1+\theta\beta z)^{-k}}{z}\mathcal{L}_{\sig}(-z)\,\mathrm dz}_{I_{1}}\IEEEnonumber\\
&&\quad\quad-\underbrace{\frac{f_{\sig}(0_{-})}{2\pi j}\int_{c-j\infty}^{c+j\infty}\frac{1-(1+\theta\beta z)^{-k}}{z}\,\mathrm dz}_{I_{2}}.\IEEEeqnarraynumspace\label{eq:intro_I}
\end{IEEEeqnarray}
The value of $c$ can be arbitrarily chosen in the interval $(-1/\theta\beta,0)$. Note that both integrands in \eqref{eq:intro_I} have a singularity at $z=z_{0}\mathdef-1/\theta\beta$. We proceed by computing the integrals $I_{1}$ and $I_{2}$ by first expressing each of them as a closed contour integral along a semi-circle to the left enclosing $z_{0}$, i.e.,
\begin{IEEEeqnarray}{rCl}
	I_{1}&=&\frac{1}{2\pi j}\left[\lim_{R\to\infty}\int_{\substack{\text{semi-circle}\\\text{of radius }R}}\frac{1-(1+\theta\beta z)^{-k}}{z}\mathcal{L}_{\sig}(-z)\,\mathrm dz\right.\IEEEnonumber\\
	&&\left.-\lim_{R\to\infty}\int_{\substack{\text{arc of}\\\text{radius }R}}\frac{1-(1+\theta\beta z)^{-k}}{z}\mathcal{L}_{\sig}(-z)\,\mathrm dz\right],\IEEEeqnarraynumspace\label{eq:arc_int}
\end{IEEEeqnarray}
with the corresponding expression for $I_{2}$. By the residue theorem \cite{lepage80}, the first integral in \eqref{eq:arc_int} is determined by the residue of the integrand at $z=z_{0}$. Using the substitution $z\to Re^{j\phi}-c$ with $\partial z/\partial\phi = jRe^{\phi}$, we write
\begin{IEEEeqnarray}{rCl}
		I_{1}&=&\text{Res}\left\{\frac{1-(1+\theta\beta z)^{-k}}{z}\lap_\sig(-z),z=z_{0}\right\}\IEEEnonumber\\
		&&-\frac{1}{2\pi }\lim_{R\to\infty}\int_{\frac{\pi}{2}}^{\frac{3}{2}\pi}\left[1-(1+\theta\beta (Re^{j\phi}-c))^{-k}\right]\,\IEEEnonumber\\
		&&\hspace{2.5cm}\times\lap_{\sig}(-(Re^{j\phi}-c))\,\mathrm d\phi.\label{eq:arc_int_tmp}
\end{IEEEeqnarray}
The integrand in \eqref{eq:arc_int_tmp} is bounded above by one, hence by the dominated convergence theorem
\begin{IEEEeqnarray}{rCl}
		&&\lim_{R\to\infty}\int_{\frac{\pi}{2}}^{\frac{3}{2}\pi}\left[1-(1+\theta\beta (Re^{j\phi}-c))^{-k}\right]\IEEEnonumber\\
		&&\hspace{3.5cm}\times\lap_{\sig}(-(Re^{j\phi}-c))\,\mathrm d\phi\IEEEnonumber\\
		&&\quad=\int_{\frac{\pi}{2}}^{\frac{3}{2}\pi}\lim_{R\to\infty}\underbrace{[1-(1+\theta\beta (Re^{j\phi}-c))^{-k}]}_{\to1}\IEEEnonumber\\
		&&\hspace{2.5cm}\times\underbrace{\lap_{\sig}(-(Re^{j\phi}-c))}_{\to f_{\sig}(0_{-})}\,\mathrm d\phi=\pi f_{\sig}(0_{-}).\IEEEeqnarraynumspace\label{eq:arc_int2}
\end{IEEEeqnarray}

The fact that $\lap_{\sig}(-(Re^{j\phi}-c))\to f_{\sig}(0_{-})$ as $R\to\infty$ uniformly for all $\phi\in[\tfrac{\pi}{2},\tfrac{3}{2}\pi]$ follows from the initial value theorem \cite{lepage80}. The residue $\text{Res}\{\lap_\sig(-z)(1-(1+\theta\beta z)^{-k})/z,z=z_{0}\}$ can be obtained by a Laurent series expansion at $z=z_{0}$ if $\lap_\sig(-z)(1-(1+\theta\beta z)^{-k})/z$ is holomorphic. To ensure holomorphy it is necessary that $k$ is integer-valued. Thus, we replace $k$ by $\tilde k=\lfloor k \rfloor$ (respectively $\tilde k=\lceil k \rceil$).
The Laurent series of $(1-(1+\theta\beta z)^{-k})/z$, now having a pole of order $\tilde k$ at $z=z_{0}$, is then given by

\begin{IEEEeqnarray}{rCl}
	\frac{1-(1+\theta\beta z)^{-\tilde k}}{z}=\sum_{\ell=-\tilde k}^{-1}(\theta\beta)^{\ell+1}(z+\tfrac{1}{\theta\beta})^{\ell}.\label{eq:gamma_laurent}
\end{IEEEeqnarray}
As for the function $\lap_\sig(-z)$, we use a Taylor expansion around the same point $z=z_{0}$, yielding

\begin{IEEEeqnarray}{rCl}
	\lap_\sig(-z)=\sum\limits_{m=0}^{\infty}\frac{\lap_\sig^{(m)}(-z)}{m!}(z+\tfrac{1}{\theta\beta})^{m}.
\end{IEEEeqnarray}

Recall that we seek the residue of $\lap_\sig(-z)(1-(1+\theta\beta z)^{-k})/z$ at $z=z_{0}$. By the Cauchy integral formula \cite{olver10}, the residue is determined by the coefficient $a_{-1}$ of the corresponding Laurent series. Thus,
\begin{IEEEeqnarray}{rCl}
	&&\text{Res}\left\{\sum_{\ell=-\tilde k}^{-1}\sum\limits_{m=0}^{\infty}(\theta\beta)^{\ell+1}\frac{\lap_\sig^{(m)}(-z)}{m!}(z+\tfrac{1}{\theta\beta})^{m+\ell},z=z_{0}\right\}\IEEEnonumber\\
	&&\quad\quad=\sum_{m=0}^{\tilde k-1}\frac{\lap_\sig^{(m)}(-z)}{m!}(\theta\beta)^{-m}.\IEEEeqnarraynumspace
\end{IEEEeqnarray}
Hence, $I_{1}=\sum_{m=0}^{\tilde k-1}\frac{\lap_\sig^{(m)}(-z)}{m!}(\theta\beta)^{-m}-\tfrac{f_{\sig}(0_{-})}{2}$. For evaluating $I_{2}$, we use the same procedure:
\begin{IEEEeqnarray}{rCl}
	I_{2}&=&\frac{f_{\sig}(0_{-})}{2\pi j}\lim_{R\to\infty}\int_{\substack{\text{semi-circle}\\\text{of radius }R}}\frac{1-(1+\theta\beta z)^{- k}}{z}\,\mathrm dz\IEEEnonumber\\
	&&-\frac{f_{\sig}(0_{-})}{2\pi j}\lim_{R\to\infty}\int_{\substack{\text{arc of}\\\text{radius }R}}\frac{1-(1+\theta\beta z)^{-k}}{z}\,\mathrm dz.\IEEEeqnarraynumspace\label{eq:arc_int3}
\end{IEEEeqnarray}
Replacing $k$ by $\tilde k$ and noting that by \eqref{eq:gamma_laurent} the residue of $(1-(1+\theta\beta z)^{-k})/z$ at $z=z_{0}$ is one, the first integral in \eqref{eq:arc_int3} is $f_{\sig}(0_{-})$. Similarly, by \eqref{eq:arc_int2}, the second integral in \eqref{eq:arc_int3} becomes $f_{\sig}(0_{-})/2$, thus $I_{2}=f_{\sig}(0_{-})/2$. Finally, plugging $I_{1}$ and $I_{2}$ back into \eqref{eq:intro_I} yields the result.\qed

\subsection{Proof of Lemma~\ref{lem:laplace_sig_bpp}}\label{ap:laplace_sig_bpp}
We write
\begin{IEEEeqnarray}{rCl}
\mathbb{E}\left[e^{-s\sig}\right]&\overset{\text{(a)}}{=}&\mathbb{E}\left[e^{-s\mathsf{g}\|\mathsf{x}\|^{-\alpha}\mathds{1}(\mathsf{g}\|\mathsf{x}\|^{-\alpha}\geq T)}\right]^{K}\IEEEnonumber\\
	&\overset{\text{(b)}}{=}&\mathbb{E}\left[\int_{0}^{D}\frac{2r}{D^2}\,e^{-s\mathsf{g}r^{-\alpha}\mathds{1}(\mathsf{g}r^{-\alpha}\geq T)}\,\mathrm dr\right]^{K}\IEEEnonumber\\
	&\overset{\text{(c)}}{=}&\mathbb{E}\left[\frac{2}{\alpha D^2}\int_{D^{-\alpha}}^{\infty}t^{-\frac{2}{\alpha}-1}\,e^{-s\mathsf{g}t\mathds{1}(t\geq T/\mathsf{g})}\,\mathrm dt\right]^{K}\IEEEnonumber\\
	&\overset{\text{(d)}}{=}&\mathbb{E}\left[\frac{-t^{-\frac{2}{\alpha}}}{D^2}\bigg\lvert_{D^{-\alpha}}^{\max\{D^{-\alpha},\frac{T}{\mathsf{g}}\}}\hspace{-.3cm}+\frac{-t^{-\frac{2}{\alpha}}e^{-s\mathsf{g}t}}{D^2}\bigg\lvert^{\infty}_{\max\{D^{-\alpha}\hspace{-.07cm},\frac{T}{\mathsf{g}}\}}\right.\IEEEnonumber\\
	&&\qquad\left.-\frac{(s\mathsf{g})^{\frac{2}{\alpha}}}{D^2}\int_{\max\{D^{-\alpha},\frac{T}{\mathsf{g}}\}}^{\infty}\hspace{-.2cm}t^{-\frac{2}{\alpha}}\,e^{-s\mathsf{g}t}\,\mathrm dt\right]^{K},\IEEEeqnarraynumspace\label{eq:lemma_last}
\end{IEEEeqnarray}
where (a) follows from the independence property of BPPs \cite{stoyan95}, (b) follows from the PDF $f_{\|\mathsf{x}\|}(r)=2r/D^2$, (c) follows from the substitution $r^{-\alpha}\to t$ and (d) follows from partial integration. Evaluating \eqref{eq:lemma_last} yields the result.\qed

\subsection{Proof of Lemma~\ref{lem:laplace_sig_ppp}}\label{ap:laplace_sig_ppp}
De-conditioning \eqref{eq:lap_bpp} on $K$, where $K=\Phi(\mathcal{C})$ is Poisson with mean $\lambda\pi D^2$, we obtain
\begin{IEEEeqnarray}{rCl}
	\lap_{\sig}(s)
		&=&e^{-\lambda\pi D^2}\sum\limits_{K=0}^{\infty}\frac{(\lambda\pi D^2)^{K}}{K!}\lap_{\sig|\Phi(\mathcal{C})=K}(s)\IEEEnonumber\\
		&=&\exp\left(-\lambda\pi D^2(1+\lap_{\sig|\Phi(\mathcal{C})=K}^{\frac{1}{K}}(s))\right),
\end{IEEEeqnarray}
where $\lap_{\sig|\Phi(\mathcal{C})=K}(s)$ is given by \eqref{eq:lap_bpp}.\qed

\subsection{Proof of Lemma~\ref{lem:der_laplace_sig}}\label{ap:der_laplace_sig}
Employing the probability generating functional for PPPs\cite{stoyan95,HaenggiBook}, the Laplace transform $\lap_{\sig}$ can be calculated as $\lap_{\sig}(s)=\exp(-2\pi\lambda\int_{0}^{D}r(1-\mathbb{E}[e^{-s\mathsf{g}r^{-\alpha}\mathds{1}(\mathsf{g}r^{-\alpha}\geq T)}])\,\mathrm dr)$. The $m$-th derivative of $\log\lap_{\sig}(-s)$ can then be calculated as 
\begin{IEEEeqnarray}{rCl}
	&&\frac{\partial^{m}\log\lap_{\sig}(-s)}{\partial s^m}\IEEEnonumber\\
	&&\quad=-2\pi\lambda\frac{\partial^{m}}{\partial s^{m}}\int_{0}^{D}r\left(1-\mathbb{E}\left[e^{s\mathsf{g}r^{-\alpha}\mathds{1}(\mathsf{g}r^{-\alpha}\geq T)}\right]\right)\mathrm dr\IEEEnonumber\\
	&&\quad\overset{\text{(a)}}{=}-2\pi\lambda\frac{\partial^{m}}{\partial s^{m}}\int_{0}^{\infty}\hspace{-.15cm}\int_{0}^{D}\hspace{-.15cm}r\,f_{\mathsf{g}}(g)\left[1-e^{sgr^{-\alpha}\mathds{1}(gr^{-\alpha}\geq T)}\right]\mathrm dr\,\mathrm dg\IEEEnonumber\\
	&&\quad\overset{\text{(b)}}{=}2\pi\lambda\int_{0}^{\infty}\int_{0}^{D}r\,f_{\mathsf{g}}(g)\frac{\partial^{m}}{\partial s^{m}}e^{sgr^{-\alpha}\mathds{1}(gr^{-\alpha}\geq T)}\mathrm dr\,\mathrm dg\IEEEnonumber\IEEEeqnarraynumspace\\
	&&\quad\overset{\text{(c)}}{=}\frac{2\pi\lambda}{\alpha}\int_{0}^{\infty}\hspace{-.15cm}f_{\mathsf{g}}(g)\,g^{m}\hspace{-.15cm}\int_{\max\{D^{-\alpha},\frac{T}{g}\}}^{\infty}\hspace{-.15cm}t^{m-\frac{2}{\alpha}-1}e^{sgt}\,\mathrm dt\,\mathrm dg
	\end{IEEEeqnarray}
for $s<0$. (a) follows from Tonelli's theorem \cite{bauer92}, (b) follows from Leibniz integration rule \cite{flanders73}, and (c) uses the substitution $r^{-\alpha}\to t$. Evaluating the inner integral and inserting the point $s=-/\theta\beta$ yields the result.\qed


\begin{thebibliography}{10}
\providecommand{\url}[1]{#1}
\csname url@samestyle\endcsname
\providecommand{\newblock}{\relax}
\providecommand{\bibinfo}[2]{#2}
\providecommand{\BIBentrySTDinterwordspacing}{\spaceskip=0pt\relax}
\providecommand{\BIBentryALTinterwordstretchfactor}{4}
\providecommand{\BIBentryALTinterwordspacing}{\spaceskip=\fontdimen2\font plus
\BIBentryALTinterwordstretchfactor\fontdimen3\font minus
  \fontdimen4\font\relax}
\providecommand{\BIBforeignlanguage}[2]{{%
\expandafter\ifx\csname l@#1\endcsname\relax
\typeout{** WARNING: IEEEtran.bst: No hyphenation pattern has been}%
\typeout{** loaded for the language `#1'. Using the pattern for}%
\typeout{** the default language instead.}%
\else
\language=\csname l@#1\endcsname
\fi
#2}}
\providecommand{\BIBdecl}{\relax}
\BIBdecl


\bibitem{3gpp_tr_36819}
3GPP, ``Coordinated multi-point operation for {LTE} physical layer aspects,''
  TR 36.819, Tech. Rep., Sep. 2011.

\bibitem{gesbert10}
D.~Gesbert \emph{et~al.}, ``Multi-cell {MIMO} cooperative networks: A new look
  at interference,'' \emph{{IEEE} J. Sel. Areas Commun.}, vol.~28, no.~9, pp.
  1380--1408, Dec. 2010.

\bibitem{simeone12}
O.~Simeone \emph{et~al.}, ``Cooperative wireless cellular systems: An
  information-theoretic view,'' \emph{Found. Trends Netw.}, vol.~8, no. 1-2,
  pp. 1--177, Aug. 2012.

\bibitem{sawahashi10}
M.~Sawahashi \emph{et~al.}, ``Coordinated multipoint transmission/reception
  techniques for {LTE}-advanced [coordinated and distributed {MIMO}],''
  \emph{IEEE Wireless Commun.}, vol.~17, no.~3, pp. 26--34, Jun. 2010.

\bibitem{irmer11}
R.~Irmer \emph{et~al.}, ``Coordinated multipoint: Concepts, performance, and
  field trial results,'' \emph{{IEEE} Commun. Mag.}, vol.~49, no.~2, pp.
  102--111, Feb. 2011.

\bibitem{ZhaChe09}
J.~Zhang \emph{et~al.}, ``Networked {MIMO} with clustered linear precoding,''
  \emph{{IEEE} Trans. Wireless Commun.}, vol.~8, no.~4, pp. 1910--21, Apr.
  2009.

\bibitem{docomo10}
H.~Taoka \emph{et~al.}, ``{MIMO and CoMP in LTE-Advanced},'' NTT Docomo, Tech.
  Rep. vol. 12 No. 2, Sep. 2010.

\bibitem{li12}
J.~Li \emph{et~al.}, ``Performance evaluation of coordinated multi-point
  transmission schemes with predicted {CSI},'' in \emph{IEEE Intl. Symposium on
  Personal Indoor and Mobile Radio Commun. (PIMRC)}, 2012, pp. 1055--1060.

\bibitem{ericsson11}
Ericsson, ``Discussions on {DL CoMP} schemes,'' 3GPP TSG-RAN WG1\#66 R1-113353,
  Tech. Rep., Oct. 2011.

\bibitem{barbieri12}
A.~Barbieri \emph{et~al.}, ``Coordinated downlink multi-point communications in
  heterogeneous cellular networks,'' in \emph{IEEE Information Theory and
  Applications Workshop (ITA)}, 2012, pp. 7--16.

\bibitem{lee12}
D.~Lee \emph{et~al.}, ``Coordinated multipoint transmission and reception in
  {LTE}-advanced: deployment scenarios and operational challenges,''
  \emph{{IEEE} Commun. Mag.}, vol.~50, no.~2, pp. 148--155, Feb. 2012.

\bibitem{morimoto06}
A.~Morimoto, K.~Higuchi, and M.~Sawahashi, ``Performance comparison between
  fast sector selection and simultaneous transmission with soft-combining for
  intra-node {B} macro diversity in downlink {OFDM} radio access,'' in
  \emph{IEEE 63rd Vehicular Technology Conference (VTC-Spring)}, vol.~1, 2006,
  pp. 157--161.

\bibitem{gosh10}
A.~Ghosh, J.~Zhang, J.~G. Andrews, and R.~Muhamed, \emph{Fundamentals of LTE},
  1st~ed.\hskip 1em plus 0.5em minus 0.4em\relax Upper Saddle River, NJ, USA:
  Prentice Hall Press, 2010.

\bibitem{lozano12}
A.~Lozano, R.~W. {Heath~Jr.}, and J.~G. Andrews, ``Fundamental limits of
  cooperation,'' \emph{{IEEE} Trans. Inf. Theory}, vol.~59, no.~9, pp.
  5213--5226, Sep. 2013.

\bibitem{tukmanov13}
A.~Tukmanov, Z.~Ding, S.~Boussakta, and A.~Jamalipour, ``On the impact of
  network geometric models on multicell cooperative communication systems,''
  \emph{IEEE Wireless Commun.}, vol.~20, no.~1, pp. 75--81, Feb. 2013.

\bibitem{stoyan95}
D.~Stoyan, W.~Kendall, and J.~Mecke, \emph{Stochastic Geometry and its
  Applications}, 2nd~ed.\hskip 1em plus 0.5em minus 0.4em\relax Wiley, 1995.

\bibitem{HaenggiBook}
M.~Haenggi, \emph{Stochastic Geometry for Wireless Networks}.\hskip 1em plus
  0.5em minus 0.4em\relax Cambridge University Press, 2012.

\bibitem{tanbourgi13_1}
R.~Tanbourgi, H.~{J{\"{a}}kel}, and F.~K. Jondral, ``Cooperative relaying in a
  {Poisson} field of interferers: A diversity order analysis,'' in \emph{IEEE
  Intl. Symposium on Inf. Theory (ISIT)}, 2013, pp. 3100--3104.

\bibitem{tanbourgi13_2}
R.~Tanbourgi, H.~S. Dhillon, J.~G. Andrews, and F.~K. Jondral, ``{Effect of
  Spatial Interference Correlation on the Performance of Maximum Ratio
  Combining},'' \emph{{IEEE} Trans. Wireless Commun.}, vol.~13, no.~6, pp.
  3307--3316, Jun. 2014.

\bibitem{dhillon12}
H.~S. Dhillon, R.~K. Ganti, F.~Baccelli, and J.~G. Andrews, ``Modeling and
  analysis of {K}-tier downlink heterogeneous cellular networks,'' \emph{{IEEE}
  J. Sel. Areas Commun.}, vol.~30, no.~3, pp. 550 -- 560, Apr. 2012.

\bibitem{marsch11}
P.~Marsch and G.~Fettweis, Eds., \emph{Coordinated Multi-Point in Mobile
  Communications}.\hskip 1em plus 0.5em minus 0.4em\relax Cambridge University
  Press, 2011.

\bibitem{keeler13}
H.~P. Keeler, B.~B{\l}aszczyszyn, and M.~K. Karray, ``{SINR}-based coverage
  probability in cellular networks under multiple connections,'' in \emph{IEEE
  Intl. Symposium on Inf. Theory (ISIT)}, 2013, pp. 1167--1171.

\bibitem{huang11}
K.~Huang and J.~Andrews, ``A stochastic-geometry approach to coverage in
  cellular networks with multi-cell cooperation,'' in \emph{IEEE Global
  Telecommunications Conference (GlobeCom)}, 2011, pp. 1--5.

\bibitem{huang12}
------, ``Characterizing multi-cell cooperation via the outage-probability
  exponent,'' in \emph{IEEE Intl. Conf. on Commun. (ICC)}, 2012, pp.
  6411--6415.

\bibitem{jung13}
S.~Y. Jung, H.-K. Lee, and S.-L. Kim, ``Worst-case user analysis in {Poisson
  Voronoi} cells,'' \emph{{IEEE} Commun. Lett.}, vol.~17, no.~8, pp.
  1580--1583, Aug. 2013.

\bibitem{baccelli13_1}
A.~Giovanidis and F.~Baccelli, ``A stochastic geometry framework for analyzing
  pairwise-cooperative cellular networks,'' \emph{ArXiv e-prints}, May 2013,
  available at \url{http://arxiv.org/abs/1305.6254}.

\bibitem{haenggi13_comp}
G.~Nigam, P.~Minero, and M.~Haenggi, ``Coordinated multipoint in heterogeneous
  networks: A stochastic geometry approach,'' in \emph{IEEE Globecom Workshop
  on Emerging Technologies for LTE-Advanced and Beyond 4G}, Dec. 2013, pp.
  145--150.

\bibitem{andrews11}
J.~Andrews, F.~Baccelli, and R.~Ganti, ``A tractable approach to coverage and
  rate in cellular networks,'' \emph{{IEEE} Trans. Commun.}, vol.~59, no.~11,
  pp. 3122 --3134, Nov. 2011.

\bibitem{blas12}
B.~B{\l}aszczyszyn and M.~Karray, ``Linear-regression estimation of the
  propagation-loss parameters using mobiles' measurements in wireless cellular
  network,'' in \emph{Intl. Symposium on Modeling and Optimization in Mobile,
  Ad Hoc and Wireless Networks (WiOpt)}, 2012, pp. 54--59.

\bibitem{heath12}
R.~W. {Heath~Jr.}, M.~Kountouris, and T.~Bai, ``Modeling heterogeneous network
  interference using {Poisson} point processes,'' \emph{{IEEE} Trans. Signal
  Process.}, vol.~61, no.~16, pp. 4114--4126, Aug. 2013.

\bibitem{singh13}
S.~Singh, H.~S. Dhillon, and J.~G. Andrews, ``Offloading in heterogeneous
  networks: Modeling, analysis, and design insights,'' \emph{{IEEE} Trans.
  Wireless Commun.}, vol.~12, no.~5, pp. 2484--2497, Dec. 2013.

\bibitem{last95}
G.~Last and A.~Brandt, \emph{Marked Point Processes on the Real Line: The
  Dynamic Approach}, ser. Probability and its Applications.\hskip 1em plus
  0.5em minus 0.4em\relax New York: Springer, 1995.

\bibitem{ganti09}
M.~Haenggi and R.~K. Ganti, ``Interference in large wireless networks,''
  \emph{Found. Trends Netw.}, vol.~3, pp. 127--248, Feb. 2009.

\bibitem{holma07}
H.~Holma and A.~Toskala, \emph{WCDMA for UMTS: HSPA Evolution and LTE}.\hskip
  1em plus 0.5em minus 0.4em\relax New York, NY, USA: John Wiley \& Sons, Inc.,
  2007.

\bibitem{lepage80}
R.~W. Lepage, \emph{Complex Variables and the Laplace Transform for
  Engineers}.\hskip 1em plus 0.5em minus 0.4em\relax Dover Publications, Inc.
  NY, 1980.

\bibitem{bauer92}
H.~Bauer, \emph{Ma\ss- und Integrationstheorie}, 2nd~ed., ser. De Gruyter
  Lehrbuch.\hskip 1em plus 0.5em minus 0.4em\relax Berlin: de Gruyter, 1992.

\bibitem{olver10}
F.~W. Olver, D.~W. Lozier, R.~F. Boisvert, and C.~W. Clark, \emph{NIST Handbook
  of Mathematical Functions}, 1st~ed.\hskip 1em plus 0.5em minus 0.4em\relax
  New York, NY, USA: Cambridge University Press, 2010.

\bibitem{feller71}
W.~Feller, \emph{An Introduction to Probability Theory and Its Applications,
  Vol. 2}, 2nd~ed.\hskip 1em plus 0.5em minus 0.4em\relax Wiley, Jan 1971.

\bibitem{bruno1857}
F.~di~Bruno, ``Note sur un nouvelle formule de calcul differentiel,'' in
  \emph{Quarterly Journal of Pure and Applied Mathematics 1}, 1857.

\bibitem{johnson07}
W.~P. Johnson, ``{The Curious History of Fa\`{a} di Bruno's Formula},''
  available at \url{http://www.maa.org/news/monthly217-234.pdf}.

\bibitem{klingenbrunn99}
T.~Klingenbrunn and P.~Mogensen, ``Modelling cross-correlated shadowing in
  network simulations,'' in \emph{IEEE Vehicular Technology Conference
  (VTC-Fall)}, vol.~3, 1999, pp. 1407--1411.

\bibitem{hassibi03}
B.~Hassibi and B.~Hochwald, ``How much training is needed in multiple-antenna
  wireless links?'' \emph{{IEEE} Trans. Inf. Theory}, vol.~49, no.~4, pp.
  951--963, Apr. 2003.

\bibitem{mondal12}
B.~Mondal \emph{et~al.}, ``Performance of downlink {CoMP} in {LTE} under
  practical constraints,'' in \emph{IEEE 23rd International Symposium on
  Personal Indoor and Mobile Radio Communications (PIMRC)}, 2012, pp.
  2049--2054.

\bibitem{zayed96}
A.~I. Zayed, \emph{Handbook of Function and Generalized Function
  Transformations}.\hskip 1em plus 0.5em minus 0.4em\relax CRC Press, 1996.

\bibitem{flanders73}
H.~Flanders, ``\BIBforeignlanguage{English}{Differentiation under the integral
  sign},'' \emph{\BIBforeignlanguage{English}{The American Mathematical
  Monthly}}, vol.~80, no.~6, pp. pp. 615--627, Dec. 1973, available at
  \url{http://www.jstor.org/stable/2319163}.

\end{thebibliography}

\IEEEtriggeratref{28}

\end{document}